%% file: test.tex
\documentclass[journal,12pt,onecolumn,draftclsnofoot]{IEEEtran}
\usepackage{amsmath,amsopn,amssymb,stackrel}
\usepackage{breqn,mathtools}
\usepackage{graphicx,xspace,color,soul}
\usepackage{epsfig}
\usepackage{longtable,multirow}
\usepackage{array,float}
\usepackage{cite,citesort}
\usepackage{url}
\usepackage{bm}
\usepackage{tabularx}

\usepackage{algorithm,algorithmicx,algpseudocode}

\usepackage{epstopdf}
\usepackage{stfloats}
\usepackage{caption}
\usepackage{subcaption}
\usepackage[switch,columnwise]{lineno}
\usepackage{color}
\usepackage[final]{changes}
\usepackage[bottom]{footmisc}

\def\h{{\mathbf h}}

\def\w{{\mathbf w}}
\def\x{{\mathbf x}}

\def\I{\mathbf{I}}

\def\bC{\mathbb{C}}

\def\cC{{\mathcal C}}

\def\cK{{\mathcal K}}

\def\cN{{\mathcal N}}

\def\cR{{\mathcal R}}

\newcommand{\norm}[1]{\|#1\|}                    
\newcommand{\Norm}[1]{\left\|#1\right\|}         

\begin{document}
\title{Energy Efficient Resource Allocation Optimization in Fog Radio Access Networks with Outdated Channel Knowledge}

\author{\IEEEauthorblockN{
Thi Ha Ly Dinh\IEEEauthorrefmark{1},
Megumi Kaneko\IEEEauthorrefmark{1}, 
Ellen Hidemi Fukuda\IEEEauthorrefmark{2},
Lila Boukhatem \IEEEauthorrefmark{3}}\\
\IEEEauthorblockA{\small{\IEEEauthorrefmark{1}National Institute of Informatics, 
2-1-2 Hitotsubashi, Chiyoda-ku, 101-8430 Tokyo, Japan\\
\IEEEauthorrefmark{2}Kyoto University, Yoshida Honmachi, 606-8501 Kyoto, Japan\\
\IEEEauthorrefmark{3}Universit\'{e} Paris-Saclay, CNRS, Laboratoire LRI, 91405 Orsay, France\\
Email: \IEEEauthorrefmark{1} \{halydinh,megkaneko\}@nii.ac.jp, \IEEEauthorrefmark{2}ellen@i.kyoto-u.ac.jp, \IEEEauthorrefmark{3}lila.boukhatem@lri.fr}}}

\maketitle

\vspace{-5mm}
\begin{abstract}
Fog Radio Access Networks (F-RAN) are gaining worldwide interests for enabling mobile edge computing for Beyond 5G. However, to realize the future real-time and delay-sensitive applications, F-RAN tailored radio resource allocation and interference management become necessary.
This work investigates user association and beamforming issues for providing energy efficient F-RANs. We formulate the energy efficiency maximization problem, where the F-RAN specific constraint to guarantee local edge processing is explicitly considered. To solve this intricate problem, we design an algorithm based on the Augmented Lagrangian (AL) method. Then, to alleviate the computational complexity, a heuristic low-complexity strategy is developed, where the tasks are split in two parts: one solving for user association and Fog Access Points (F-AP)  activation in a centralized manner at the cloud, based on global but outdated user Channel State Information (CSI) to account for fronthaul delays, and the second solving for beamforming in a distributed manner at each active F-AP based on perfect but local CSIs. Simulation results show that the proposed heuristic method achieves an appreciable performance level as compared to the AL-based method, while largely outperforming the energy efficiency of the baseline F-RAN scheme and limiting the sum-rate degradation compared to the optimized sum-rate maximization algorithm\footnote{Part of this paper has been presented at IEEE CCNC 2019 \cite{dinh2019energy}}.
\end{abstract}

\IEEEpeerreviewmaketitle

\section{Introduction}
\label{sec:intro}
\input{intro}

\section{Related Work}
\label{sec:related}
\input{relatedworks}

\section{System and Energy Consumption Models}
\label{sec:system}
\input{system}

\section{Optimization Problem}
\label{sec:formulation}
\input{formulation}

\section{Proposed Algorithms}
\label{sec:proposed}
\input{proposed}

\section{Numerical Results}
\label{sec:results}
\input{results}

\section{Conclusion}
\label{sec:conclude}
\input{conclude}

\section*{Acknowledgments}
This work was funded in part by the Grant-in-Aid for Scientific Research (Kakenhi) no. 17K06453 from the Ministry of Education, Science, Sports and Culture of Japan.

\bibliographystyle{IEEEtran}
\bibliography{ref2}
\end{document}

%% file: intro.tex
Given the ever-increasing number of wireless subscribers, predicted to exceed nine billion in 2022 \cite{cisco2019online}, along with the expansion of Internet of Things (IoT) communications, the volume of mobile data traffic will experience an exponential growth in the near future. 
As a result, current wireless systems will soon become unable to cope with the more and more stringent Quality of Service (QoS) levels required by the future wireless services and applications, such as Extreme Reality (XR), remote medical care or self-driving \cite{6Gpaper}.                                                                                                                                                         Hence, next generation mobile systems such as Beyond 5G and 6G will be facing tremendous technological challenges, as they are expected to jointly achieve higher capacity, lower latency, massive support of IoT devices, lower network costs, as well as higher energy efficiency.

\begin{figure}[t]
\centering
\begin{minipage}{.45\textwidth}
  \centering
  \includegraphics[scale=0.5]{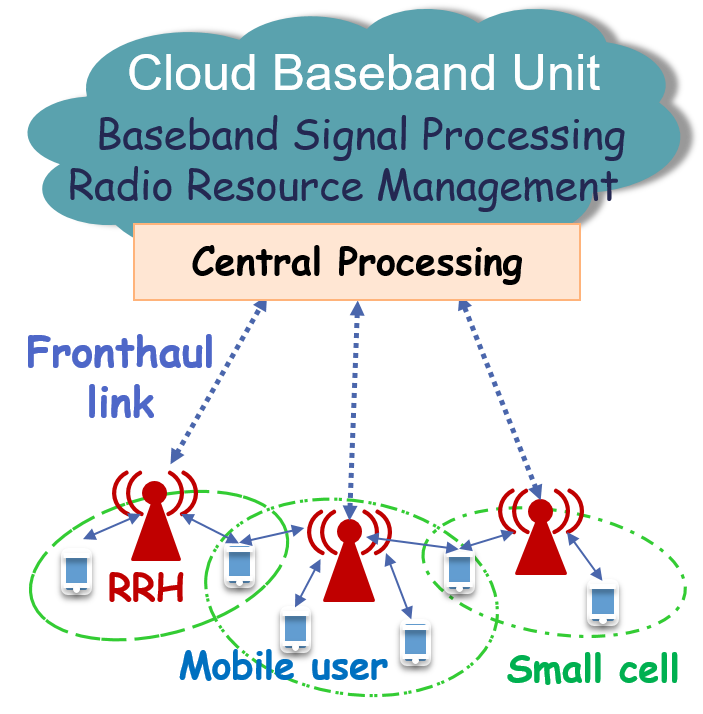}
  \captionof{figure}{C-RAN architecture}
  \label{fig:CRAN}
\end{minipage}%
\begin{minipage}{.1\textwidth}
  \centering
  \includegraphics[scale=0.35]{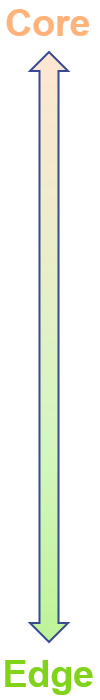}
\end{minipage}%
\begin{minipage}{.45\textwidth}
  \centering
  \includegraphics[scale=0.5]{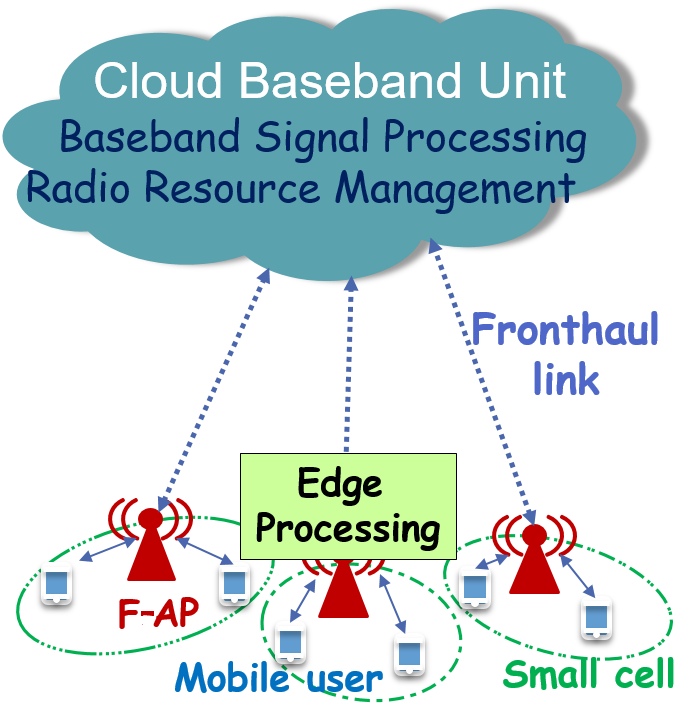}
  \captionof{figure}{F-RAN architecture}
  \label{fig:FogRAN}
\end{minipage}
\end{figure}

Towards this goal, the Cloud Radio Access Network (C-RAN) architecture has been proposed as a powerful candidate for supporting 5G and Beyond systems \cite{checko2014cloud}. This architecture enables
the control of a large number of small cells by powerful cloud-centralized processors, while deploying low cost/low power Access Points (APs) for radio access. 
In this architecture, multiple APs, referred as Remote Radio Heads (RRHs), are connected to the cloud Baseband Unit (BBU) pool through wireless or wired fronthaul links, as depicted in Fig. \ref{fig:CRAN}. The baseband signals of all users are sent via these fronthaul links to the BBU pool which then performs all tasks such as joint signal processing and radio resource management \cite{rost_cloud_2014}. Although this architecture enables global network optimization, its major drawback lies in the  high latencies that users may experience due to the transport delays incurred by these capacity-limited fronthaul links \cite{rost_cloud_2014}. Also due to this induced delays, only outdated, and hence imperfect Channel State Information (CSI) knowledge of the link qualities between APs and users will be available at the cloud BBUs, reducing the performance of resource allocation schemes \cite{wang_robust_2017}.

To overcome these high latency and fronthaul burden issues, the Fog Radio Access Network (F-RAN) architecture, depicted in Fig. \ref{fig:FogRAN},  has recently emerged. By moving the cloud and network intelligence towards the network edges, F-RANs are viewed as the enablers of the Mobile Edge Computing (MEC) paradigm \cite{shih_enabling_2017}. Unlike RRHs in C-RAN, Fog Access Points (F-AP) are now equipped with limited cloud computing capabilities, enabling them to perform partial signal processing and radio resource management tasks. By moving part of the cloud intelligence towards the network edges, F-RAN significantly alleviates the fronthaul burdens of C-RAN, thereby decreasing the delays experienced by end-users and enabling real-time processing of QoS-demanding applications. In particular, radio resource allocation tasks can be implemented with partial but perfect CSI at each F-AP.  Another specificity of F-RAN inherent to its architecture is that, each user may be associated to at most one F-AP at a time, namely to a local F-AP that processes the user's application as explained in  \cite{shih_enabling_2017}.  In the sequel, this constraint, referred to as the F-RAN specific local processing constraint, will be shown to further increase the difficulty of the considered resource allocation optimization problem.  

Until now, many recent studies for enhancing data rate, spectral and energy efficiency in C-RANs have gained deep interests from both academia and industry \cite{dai_sparse_2014,dai_backhaul-aware_2015,huang_joint_2016,awais_efficient_2017,thang_performance-cost_2017,ha_advance_2018,peng2016energy,dai_energy_2016,zuo_energy_2016}, while fewer works have dealt with the specific constraints of F-RANs. Moreover, most of these works have assumed the knowledge of perfect CSI at the cloud BBU, which is not a viable assumption due to the aforementioned transport delays on fronthaul links.  In addition, among the different performance criteria for assessing next generation wireless systems, energy consumption is regarded as one of the major concerns, so that the energy efficiency (EE) metric has been introduced as a new important factor for designing green wireless communication systems \cite{feng2012survey,wu_overview_2017}. The energy efficiency optimization problem was studied for a F-RAN in \cite{yan2018cost,na_energy_2018}, however, while \cite{yan2018cost} proposed a fully centralized algorithm using global and perfect CSI knowledge at the cloud, the proposed distributed resource allocation method in \cite{na_energy_2018} was completely decoupled from the cloud centralized computing capabilities, thereby degrading its achievable performance. 

Therefore, in this work, we aim at devising energy-efficient wireless resource allocation and interference mitigation methods in F-RANs, with a focus on user association and beamforming issues under the realistic constraint of outdated CSI knowledge at the cloud BBU pool. 

\noindent Firstly, the considered energy-efficiency optimization problem is cast as a mixed continuous and discrete problem over two variables, namely the beamforming weights and the set of active F-APs. It is worthwhile noting that, unlike in our preliminary work \cite{dinh2019energy}, power consumption models for both active and idle F-APs are accounted for, giving rise to a more realistic model, as well as a different mathematical problem formulation. This problem is mathematically  intractable as such, owing to the discrete F-RAN specific local processing constraint as well as the fronthaul constraints. To handle this, we first relax the F-RAN specific discrete constraint into a differentiable one, and then propose, unlike in \cite{dinh2019energy}, an algorithm based on the Augmented Lagrangian (AL) method. However, the high computational complexity of this algorithm makes it difficult to apply in large-scale networks. Therefore, we next propose a heuristic low-complexity strategy taking advantage of both the centralized processing capabilities of the cloud BBU and the distributed computing properties of F-APs. Namely, user scheduling and activation of energy-efficient F-APs are optimized through an equivalent sum-rate maximization problem in a centralized manner using global but outdated CSI knowledge. Next, the beamforming vectors are optimized in a distributed way using local but perfect CSI knowledge at each active F-AP.  
Computer simulation results show that our proposed methods outperform baseline schemes in terms of global energy efficiency. In particular, the heuristic method is able to achieve a good performance compared to the AL-based algorithm, while greatly enhancing the global energy efficiency of the baseline. Furthermore, the sum-rate degradation of the proposed heuristic method as compared to the reference sum-rate maximization algorithm is also shown to be limited.

To summarize, our main contributions in this work are listed as follows.
\begin{enumerate}
\item The energy efficiency optimization problem in terms of user association and beamforming for F-RAN is mathematically formulated under fronthaul, power and F-RAN specific edge processing constraints. This formulation gives rise to a non-convex mixed-integer optimization problem which is intractable as such. 
\item An AL-based EE maximization algorithm is designed to solve the formulated problem. By reformulating the initial problem into a smooth one, this AL-based method is performed centrally at the cloud BBUs assuming global and perfect CSI among all F-APs and users.
\item Given the high computational complexity required by the AL-based solution, we propose a low-complexity heuristic EE algorithm that takes advantage of the F-RAN architecture to efficiently decouple user association and beamforming tasks. Namely, user association is solved in a centralized manner at the cloud BBUs with global but outdated CSI, while beamforming is implemented distributively by each active F-AP with local but perfect CSI. 
\item Computer simulations under different network settings assess the performance improvements achieved by the proposed algorithms against baseline methods, one aiming at energy-efficiency maximization and one reference sum-rate maximizing method for F-RAN. 
\end{enumerate}

The remainder of this paper is organized as follows. In section II, we describe the related works. Next, system and energy consumption models are given in section III. Then, the energy-efficiency optimization problem in F-RAN is formulated in Section IV. In section V, our proposed algorithms are presented in details. Simulation results are discussed in section VI. Finally, section VII concludes the paper and gives directions for future works.

%% file: relatedworks.tex
In the context of C-RANs, radio resource allocation and interference management issues, including user association and beamforming, have been extensively addressed in the literature so far \cite{dai_sparse_2014,dai_backhaul-aware_2015,huang_joint_2016,awais_efficient_2017,thang_performance-cost_2017,ha_advance_2018}.   
The problem of energy efficiency maximization in a C-RAN was also considered assuming an optimal network-wide cooperative beamforming design \cite{peng2016energy}, while different heuristic user scheduling and beamforming methods were proposed \cite{dai_energy_2016,zuo_energy_2016}. 
Considering a C-RAN system equipped with cache-enabled Unmanned Aerial Vehicles (UAVs), \cite{chen2017caching} investigated the issue of joint user-UAV associations and contents caching at UAVs such that the users' quality of experience (QoE) is maximized while minimizing the transmit power between UAVs and users.

However, in the case of F-RANs, the amount of literature regarding radio resource allocation and interference management is comparatively scarce, despite the need for the F-RAN specific constraint of local processing \cite{tandon2016cloud}. Indeed,  most studies have focused on caching strategies and latency reduction as in \cite{tandon2016cloud,park2016jointLatency,shih_enabling_2017}. Along this line, \cite{park2016joint} investigated the issue of delivery rate maximization, i.e., optimizing the minimum-user rate under fronthaul capacity and F-AP power constraints. By taking into account the fronthaul cost caused by fetching contents missed in the cache, \cite{sun2016user} introduced a distributed cluster formation to maximize the whole system throughput. Under the impact of node positions, cache size and fronthaul delay costs,  \cite{yan2017evolutionary} proposed a dynamic user mode selection where each user can choose to associate with a F-AP or with another user, in order to achieve optimal payoffs for both F-APs and device-to-device users. 

Regarding joint beamforming and user association, \cite{chen2016backhaul} aimed at power consumption minimization, while guaranteeing the QoS at user sides and balancing the backhaul traffic. Namely, in their study, users requesting the same file are grouped together, and each user group is served by a cluster of F-APs. Furthermore, \cite{randrianantenaina2019interference} proposed user association, power allocation and NOMA power-split optimization algorithms for the maximizing the weighted sum-rate of a NOMA-based F-RAN. However, perfect CSI knowledge for all users is assumed at the cloud, which is not realistic since only outdated CSI will be available at the cloud BBU pool, owing to the aforementioned transport delays.
Considering this, \cite{ourpaper,kaneko2019user} took into account the impact of imperfect CSI knowledge at the cloud while handling the issue of user scheduling and beamforming in F-RANs. However, only weighted sum-rate maximization was considered.

Concerning energy efficiency, \cite{mebrek2017efficient} presented a genetic algorithm to assign users to F-APs for balancing energy consumption and delay without usage of information at the cloud. On the contrary, \cite{yan2018cost} completely handles the radio resource allocation and caching at the centralized cloud with the ideal assumption of perfect global CSI knowledge among all F-APs and users. Also with the same assumption and in a totally centralized manner, \cite{sun2019deep} investigated the issue of mode selection (C-RAN or device-to-device) and resource management in F-RAN, but without ensuring the F-RAN specific local processing constraint. To save the power of the whole system, \cite{sun2019deep} proposed to perform precoding for UEs in C-RAN mode first, then to make use of a Deep Q-network to select the on/off state for the cloud processor and user communication mode based on the transmitter cache state. Besides, \cite{na_energy_2018} proposed two heuristic algorithms to activate RRHs/F-APs for energy efficiency improvement: one that is cloud centralized and does not take advantage of F-APs' computing ability, and the second that is only carried out at each F-AP in a fully distributed manner without making any use of the cloud.

In conclusion, to the best of our knowledge, there have been very few works tackling the design of user association and beamforming for global energy efficiency optimization, under the specific constraints imposed by the F-RAN architecture and  by the impairments stemming from  outdated CSI knowledge at the cloud BBU pool, which will be the object of our work.

%% file: system.tex
\subsection{System model}
\label{systemModel}
\begin{figure}[t!]
\centering
\includegraphics[scale=0.27]{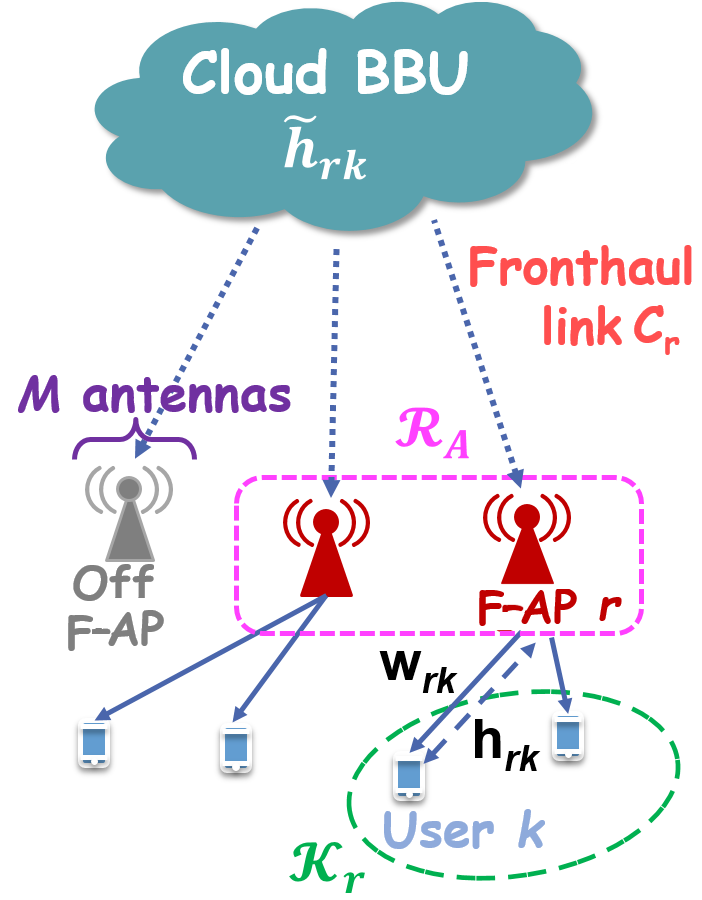}
\caption{F-RAN system model}
\label{fig:FRAN}
\end{figure}

We focus on the downlink transmissions of the F-RAN depicted in Fig. \ref{fig:FRAN}, which is composed of a cloud BBU pool that controls a set $\cR$ of  macro and pico F-APs, referred to as RRHs in conventional C-RANs, with cardinality $R=|\cR|$. Each F-AP $r$ is equipped with $M_r$ antennas and is connected to the BBU pool through a wired optical fronthaul link with capacity $C_r$. We assume $K$ single-antenna mobile users in the network. 

Denoting the beamforming vector from F-AP $r$ to user $k$ by $\w_{rk} \in \bC^{M_r \times 1}$, the beamforming vector from all F-APs to user $k$ is constructed by concatenating all these vectors into vector $$\w_k = [\w_{1k}^H,\w_{2k}^H, \cdots, \w_{Rk}^H]^H \in \bC^{M \times 1},$$
where $M = \sum_{r\in \cR}M_r$ is the sum of all $R$ F-APs' transmit antennas.
In the same way, the channel vector between F-AP $r$ and user $k$ is denoted by $\h_{rk} \in \bC^{M_r \times 1}$ while the global channel between all F-APs and user $k$ is represented by the concatenated vector $$\h_k = [\h_{1k}^H,\h_{2k}^H,\cdots,\h_{Rk}^H]^H \in \bC^{M \times 1}.$$ 

As in \cite{dai_backhaul-aware_2015}, given a transmit message $s_k$ that is independently generated from a signal constellation with zero mean and unit variance, the signal $y_k$ received by the user $k$ is expressed as
\begin{equation}
	y_k = \h_k^H\w_{k}s_{k} + \h_k^H\sum_{{\substack{k' \in \cK\\ k' \neq k}}}\w_{k'}s_{k'} + n_k,
\label{eq:ReceivedSignal}
\end{equation}
where $n_k \sim \cC\cN(0, \sigma_n^2)$ is the Additive white Gauss noise (AWGN) at user $k$. In (\ref{eq:ReceivedSignal}), the first term represents the desired signal while the second term is the interference caused by the other user's signals. 

Next, the set of all $K$ mobile users in the network is denoted as $\cK$, and $\cK_r$ represents the set of users served by F-AP $r$ with cardinality $K_r = |\cK_r|$. The achievable data rate $R_k$ at user $k$ is given by the Shannon capacity,
\begin{equation}
\label{eq:Rate}
R_k = \log(1 + \gamma_k).
\end{equation}
By setting $\mathbf{E}[|s_k|^2] = 1 $ and $\mathbf{E}[|n_k|^2] = \sigma_n^2$, the Signal to Interference-plus-Noise Ratio (SINR) $\gamma_k$ of user $k$, i.e., the ratio between the desired signal power and the interference signal plus noise power, is written as
\begin{equation}
\label{eq:SINR}
 \gamma_k = \frac{\left|\h_k^H\w_k \right|^2}{\sum\limits_{\substack{k' \in \cK \\ k' \neq k}}|\h_k^H\w_{k'}|^2 + \sigma_n^2}.
\end{equation}

In the considered F-RAN, the cloud BBU pool  has global CSI knowledge, namely the $\bm{h}_{rk}$ vectors over all F-APs $r$ and users $k$. However, due to the transport delays on fronthaul links as mentioned above, this global CSI becomes outdated, and hence is denoted $\tilde{\h}_k = [\tilde{\h}_{1k}^H,\tilde{\h}_{2k}^H,\cdots,\tilde{\h}_{Rk}^H]^H \in \bC^{M \times 1}$. 
Based on \cite{wang_robust_2017,ourpaper}, imperfect channel vector $\tilde{\h}_{rk}$ is assumed as  
\begin{equation}
\label{eq:outdateCSI}
\tilde{\h}_{rk} = \h_{rk} + \bm{e}_{rk},
\end{equation}
with the stochastic error $\bm{e}_{rk} \sim \cC\cN(0, \sigma_e^2\I_{M_r})$ where $\sigma_e^2$ denotes the error variance and $\I_{M_r}$ is the $M_r \times M_r$ identity matrix. By contrast, each F-AP possesses perfect CSI knowledge for its locally associated users, however it is oblivious to the CSIs of other users. Let us note that, the impact of such outdated CSI knowledge at the BBU pool will be accounted for in the proposed algorithm design\footnote{In order to assess the actual impact of outdated CSI, in this work, we do not assume a specific statistical channel model over multiple frames. However, such a model may be well incorporated in our method, in particular under high user mobility as in \cite{ha2019adaptive}. These issues will be considered in the follow-up work.}, as in \cite{ourpaper,kaneko2019user}.

\vspace{-3mm}
\subsection{Energy consumption model}
\label{energyModel}
\begin{figure}[t]
\centering
\includegraphics[scale=0.37]{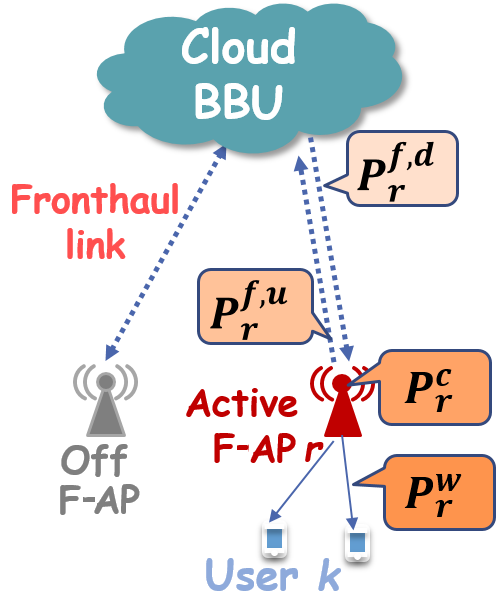}
\caption{Power consumption model}
\label{fig:power}
\end{figure}

In this section, we introduce the power consumption model for F-RANs depicted in Fig. \ref{fig:power}. Hereafter, the model of \cite{na_energy_2018} will be globally followed in order to define the power consumption $P_r$ related to F-AP $r$. Parameter $P_r$ is composed of:   
\begin{itemize}
	\item  $P_r^{f,u}$: consumed power for user CSIs transmission from F-AP $r$ to the BBU pool through the fronthaul link,  
	\item  $P_r^c$: consumed power for the circuit at F-AP $r$,
	\item  $P_r^{f,d}$: consumed power on the fronthaul link of F-AP $r$ by the downlink transmissions from the cloud BBU, 
	\item  $P_r^{w}$: consumed power for downlink wireless transmissions to F-AP $r$'s associated users in set $\cK_r$. 
\end{itemize}

\begin{table}[h!]
\caption{Parameter description of the power consumption model}
\centering
\begin{tabularx}{0.7\textwidth}{c | X }
\hline \hline
\textbf{Parameter} & \textbf{Description} \\ \hline
$P_{\textrm{fix},r}$ & Traffic-independent fixed power consumption of $r$-th fronthaul\\ \hline
$\beta$ & Redundancy in the fronthaul transport interface\\ \hline
$b_{IQ}$ & Number of IQ sampling bits, where IQ samples are digitized and encapsulated by a fronthaul transport interface \cite{park2015large} \\ \hline
$f_{\textrm{pre}}$ & Frequency of updating the precoder\\ \hline
$P_{\textrm{td},r}$ & Traffic-dependent power \\ \hline
$\beta_r$ & Efficiency of power amplifier\\ \hline
$\rho_d$ & Normalized downlink transmit power \\ \hline
$N_0$ & Noise power \\ \hline
$P_{\textrm{ic},r}$ & Power for antenna circuit components of F-AP \\ \hline
$B$ & Wireless channel bandwidth\\ 
\hline \hline
\end{tabularx}
\label{tb:parameter}
\end{table}

Each of the components above is calculated as follows, given the parameters described in Table \ref{tb:parameter},
\begin{align}
P_r^{f,u} &= P_{\textrm{fix},r} + \beta M_r|K_r|b_{IQ}f_{\textrm{pre}}P_{\textrm{td},r}, \label{eq1} \\[2mm]
P_r^c &= \frac{1}{\beta_r}\rho_dN_0 + M_rP_{\textrm{ic},r}, \label{eq2} \\[2mm]
P_r^{f,d}(\w_{rk}, \cR_A) &= P_{\textrm{fix},r} + B\tau(\w_{rk}, \cR_A) P_{\textrm{td},r}, \label{eq3} \\[2mm]
P_r^{w}(\w_{rk}) &= \sum_{k \in \cK_r}||\w_{rk}||_2^2. \label{eq4}
\end{align}
In (\ref{eq3}), the spectral efficiency $\tau$ is defined as the sum of user rates, which is a function of the beamforming vectors $\w_{rk}$ and the set of active F-APs $\cR_A$, 
\begin{equation}
\tau(\w_{rk}, \cR_A) = \sum_{r \in \cR_A}\sum_{k \in \cK_r}R_k.
\label{eq:tau}
\end{equation}
With the calculation of components above, $P_r^{f,u}$ and $P_r^c$ are considered as constants due to their independence on beamforming vectors $\w_{rk}$ and set of active F-APs $\cR_A$.

The power consumed by the set of active F-APs $\cR_A$, in each frame, is given as
\begin{equation}
\label{eq:PF}
\begin{array}{lcl}
P^{\text{on}}(\w_{rk}, \cR_A)&=&\sum\limits_{r \in \cR_A}P_r^{\text{on}}(\w_{rk})  \\
&=&\sum\limits_{r \in \cR_A} \left(P_r^{f,u} + P_r^c + P_r^{w}(\w_{rk})\right) + P_r^{f,d}(\w_{rk}, \cR_A).
\end{array} 
\end{equation}
In the proposed algorithms, in order to improve energy efficiency, some F-APs can be switched off in certain frames. F-APs that are switched-off are said to be in an idle state and hence only consume power
\begin{equation}
P^{\text{off}}(\cR_A) = \sum_{r \in \cR \backslash \cR_A}P_r^{\text{off}} = \sum_{r \in \cR \backslash \cR_A} P_r^{f,u} + P_r^c.
\end{equation}

Finally, the total consumed power $P$ of the system is determined by
\begin{equation}
\begin{array}{lcl}
P(\w_{rk}, \cR_A) &=& P^{\text{on}}(\w_{rk}, \cR_A) +  P^{\text{off}}(\cR_A) \\
&=& \sum\limits_{r \in \cR_A} \left(P_r^{f,u} + P_r^c + P_r^{w}(\w_{rk})\right) + P_r^{f,d}(\w_{rk}, \cR_A) + \sum\limits_{r \in \cR \backslash \cR_A} (P_r^{f,u} + P_r^c).
\label{eq:power}
\end{array}
\end{equation}

%% file: formulation.tex
\label{formulation}
Based on the system and power consumption models above, we can now express the mathematical formulation of the energy efficiency optimization problem of user association and beamforming under fronthaul, power and F-RAN specific edge processing constraints.

First of all, the global energy efficiency $\eta$ is defined below as the ratio between the global throughput $B\tau$, where $B$ denotes the downlink transmission bandwidth, and its total power consumption $P$ \cite{na_energy_2018},
\begin{equation}
\label{eqEE}
\eta(\w_{rk}, \cR_A) = \frac{B\tau(\w_{rk}, \cR_A)}{P(\w_{rk}, \cR_A)}.
\end{equation}

Substituting $\tau(\w_{rk}, \cR_A)$ and $ P(\w_{rk}, \cR_A)$ by Eqs. (\ref{eq:tau}) and (\ref{eq:power}), we can express the energy efficiency optimization problem as
\begin{IEEEeqnarray}{c}
\label{eq:EEproblem} \IEEEyessubnumber*
\mkern-28mu\max_{{\w}_{rk}, \cR_A} \eta(\w_{rk}, \cR_A) =  \frac{B\sum_{k \in \cK}R_k}{\sum\limits_{r \in \cR \backslash \cR_A} (P_r^{f,u} + P_r^c) + \sum\limits_{r \in \cR_A}\left(P_r^{f,u} + P_r^c + P_r^{w}(\w_{rk})\right) + P_r^{f,d}(\w_{rk},\cR_A)}, \nonumber \IEEEyesnumber\\
 \quad \quad  \quad \quad  \quad \quad \mbox{s.t., } \quad \quad P_r^{w}(\w_{rk}) = \sum_{k \in \cK_r}||\w_{rk}||_2^2  \leq  P_r^{\max}, \hfill \forall r \in \cR_A, \quad \quad \quad \quad \quad \quad \quad \label{eq:PrConstraint} \IEEEyessubnumber \\[2mm] 
 \quad \quad  \quad \quad  \quad \quad   \quad \quad  \quad \quad  \quad \quad  \quad \quad  \quad \quad \quad \  \sum_{k \in \cK_r}R_k   \leq  C_r, \hfill \forall r \in \cR_A, \quad \quad \quad \quad \quad \quad \quad \label{eq:CrConstraint} \IEEEyessubnumber \\[2mm]
 \quad \quad  \quad \quad  \quad \quad   \quad \quad  \quad \quad  \quad \quad \quad \quad \ \sum_{r \in \cR_A}\Big\|||\w_{rk}||_2^2\Big\|_0   \leq 1, \hfill \forall k \in \cK. \quad \quad \quad \quad \quad \quad \quad
	\label{eq:FogConstraint}\IEEEyessubnumber
\end{IEEEeqnarray}
The first constraint is given by the transmit power budget $P_r^{\max}$ of each F-AP $r$. Constraint (\ref{eq:CrConstraint}) sets the fronthaul capacity limitation. Since $\w_{rk}$ is non-zero if and only if user $k$ is associated to F-AP $r$, while $||\x||_0$ is the number of non-zero elements of $\x$, the last inequality expresses the F-RAN specific constraint whereby each user can be served by at most one F-AP as aforementioned. This is to ensure the F-RAN mandatory local edge operation as explained in \cite{peng_fog-computing-based_2016}. 

Problem (\ref{eq:EEproblem}) is a non-convex optimization over variables $\w_{rk}$ and $\cR_A$, which is difficult to solve. In particular, the discrete constraint (\ref{eq:FogConstraint}) is on both variables and is even more challenging than in the weighted sum-rate maximization problem for F-RANs of \cite{ourpaper,kaneko2019user}\footnote{Given the difficulty of this problem, minimum QoS constraints as in \cite{d2019power,aqeeli2017power} have not been included. However, Section VI.C shows that the proposed method still improves the achievable rate of worst CSI users compared to baseline.}. To handle this issue, we first approximate the constraint (\ref{eq:FogConstraint}) and then propose a solution by an approach based on Augmented Lagrangian method. However, this algorithm requires a huge computation complexity, and therefore we also propose a heuristic strategy composed of two phases for treating each variable separately, while accounting for the CSIs' imperfectness at cloud BBUs as well as the limited-complexity requirements of F-APs' operations. The details of the two proposed approaches are given in the next section.

%% file: proposed.tex

\subsection{Augmented Lagrangian-based Algorithm for EE-Maximization}
\label{optimization}
As the $\ell_0$-norm in constraint (\ref{eq:FogConstraint}) is non-differentiable, problem~\eqref{eq:EEproblem} is nonsmooth. Thus, we instead consider twice
continuously differentiable approximations of the $\ell_0$-norm for this constraint. 
One of the simplest approximation of $\norm{z}_0$ with $z \in \bC^L$ is defined as
\begin{equation}
  \label{eq:l0_aprox_1}
  z \mapsto \sum_{i=1}^L \frac{z_i}{z_i + \delta_i},
\end{equation}
with sufficiently small $\delta_i > 0$ for all $i=1,\dots,L$. It can be easily observed that \eqref{eq:l0_aprox_1} is a well-defined function and twice continuously
differentiable. Since in constraint \eqref{eq:FogConstraint}, we only require the $\ell_0$-norm of a scalar term, i.e., $L = 1$, we define the following functions $\Phi^\ell \colon \bC^\ell \to \mathbb{R}$  as
\begin{equation}
\Phi^\ell(y) := \frac{\norm{y}_2^2}{\norm{y}_2^2 + \delta},
\label{eq:appx}
\end{equation}
where $\delta > 0$ is sufficiently small. Note that $\Phi^\ell$ is also twice continuously differentiable because $\norm{\cdot}_2^2$ and \eqref{eq:l0_aprox_1}
are twice continuously differentiable. Moreover, $\Phi^\ell$ approximates the $\ell_0$-norm of the square of the $\ell_2$-norm, i.e., $\Norm{\norm{\cdot}_2^2}_0$.\\[2mm]

In addition, as mentioned in Section \ref{formulation}, the beamforming vector $\w_{rk}$ is non-zero if and only if user $k$ is associated to F-AP $r$, thus we can write $\w_{rk} = \mathbf{0}, \forall k \in \cK$, if F-AP $r \notin \cR_A$. Therefore,  constraint (\ref{eq:FogConstraint}) can be approximated by
\begin{equation}
\sum_{r \in \cR_A}\Big\|||\w_{rk}||_2^2\Big\|_0 \approx \sum_{r \in \cR_A}\Phi^{M_r}(\w_{rk}) = \sum_{r \in \cR}\Phi^{M_r}(\w_{rk}) \leq 1, \quad \forall k \in \cK.
\end{equation}

Similarly, constraints (\ref{eq:PrConstraint}), (\ref{eq:CrConstraint}) can be respectively rewritten as
\begin{equation}
P_r^{w}(\w_{rk}) = \sum_{k \in \cK_r}||\w_{rk}||_2^2 = \sum_{k \in \cK}||\w_{rk}||_2^2 \leq P_r^{\max}, \quad \forall r \in \cR, \quad \text{and}
\end{equation}
\begin{equation}
\sum_{k \in \cK_r}R_k \approx \sum_{k \in \cK_r}R_k \Phi^{M_r}(\w_{rk}) = \sum_{k \in \cK}R_k \Phi^{M_r}(\w_{rk})  \leq C_r, \quad \forall r \in \cR.
\end{equation}

Finally, for convenience, we represent the total power consumption as
\begin{equation}
\begin{array}{lcl}
P(\w_{rk}, \cR_A) &=&\sum\limits_{r \in \cR \backslash \cR_A} (P_r^{f,u} + P_r^c) + \sum\limits_{r \in \cR_A} (P_r^{f,u} + P_r^c + P_r^{w}(\w_{rk}))+ P_r^{f,d}(\w_{rk},\cR_A) \\
&=& \sum\limits_{r \in \cR}(P_r^{f,u} + P_r^c) + \left(P_r^{f,d}(\w_{rk}) + P_r^w(\w_{rk})\right)\Phi^{KM_r}(\w_r),
\end{array}
\end{equation}
where $\w_r = [\w_{r1}^H,\w_{r2}^H, \cdots, \w_{rK}^H]^H \in \bC^{KM_r \times 1}$ is the beamforming vector from F-AP $r$ to all users.

Now, we can reformulate \eqref{eq:EEproblem} into the smooth optimization problem as follows:
\begin{IEEEeqnarray}{rl}
\label{eq:reform} \IEEEyessubnumber*
\min_{{\w}_{rk}} \quad &  \frac{-B\sum_{k \in \cK}R_k}{\sum\limits_{r \in \cR}(P_r^{f,u} + P_r^c) + (P_r^{f,d} + P_r^w)\Phi^{KM_r}(\w_r)},\nonumber \IEEEyesnumber\\
\mbox{s.t.,} \quad &	P_r^{w}(\w_{rk}) = \sum_{k \in \cK}||\w_{rk}||_2^2  \leq  P_r^{\max}, \hfill \forall r \in \cR, \label{eq:PrConstraintRe} \IEEEyessubnumber \\[1.5mm]
&\quad \quad \quad \ \sum_{k \in \cK}R_k \Phi^{M_r}(\w_{rk}) \leq  C_r, \hfill  \forall r \in \cR,\label{eq:CrConstraintRe} \IEEEyessubnumber \\[1.5mm]
&\quad \quad \quad \quad \quad \sum_{r \in \cR}\Phi^{M_r}(\w_{rk}) \leq 1, \hfill \forall k \in \cK.
	\label{eq:FogConstraintRe}\IEEEyessubnumber
\end{IEEEeqnarray}

In this reformulation, we solve for the overall beamforming vector $\w$ which contains all beamforming vectors $\w_{rk}$ from F-APs to users. Then, based on the obtained solution, we can retrieve the optimal set of active F-APs $\cR_A$ as well as their associated users from the condition: $k \in \cK_r $ if and only if $\ {\w}_{rk} \neq \bm{0}$. \\[1.5mm]

We observe that problem (\ref{eq:reform}) has non-linear non-convex objective function and constraints. Therefore, we propose to solve it by means of the Augmented Lagrangian (AL) method \cite{birgin2008improving}. This algorithm seeks the solution of the original constrained problem by solving a sequence of unconstrained subproblems. Namely, for a given parameter $\rho > 0$, called the penalty
parameter, the \emph{Augmented Lagrangian function} associated to \eqref{eq:reform} is given by
\begin{eqnarray*}
  L_\rho(\w,\mu_1,\mu_2,\mu_3) & := & f(\w) + \frac{\rho}{2} \sum_{j=1,2} \:
  \sum_{r \in \cR} \max \left\{ g_{jr}(\w) + \frac{\mu_{jr}}{\rho} ,0 \right\}^2 \\
  & & \hfill + \frac{\rho}{2} \sum_{k \in \cK} \max \left\{ g_{3k}(\w) + \frac{\mu_{3k}}{\rho} ,0 \right\}^2,
  \label{eq:subprobLagran}
\end{eqnarray*}
where $\mu_{1r},\mu_{2r} \in \mathbb{R}$, $\forall r \in \cR$, and $\mu_{3k} \in \mathbb{R}$, $\forall k \in \cK$, are the Lagrange multipliers, and

\begin{equation*}
\begin{array}{llll}
  f(\w) & := & \displaystyle{\frac{-B \displaystyle{\sum_{k \in \cK} R_k}}{
      \displaystyle{\sum_{r \in \cR} P_r^{f,u} + P_r^c + (P_r^{f,d} + P_r^w) 
          \Phi^{KM_r}(\w_r) }}}, \\
  g_{1r}(\w) & := & \displaystyle{\sum_{k \in \cK_r} \norm{\w_{rk}}^2_2 - P_r^{\max}}
  &\forall r \in \cR, \\
  g_{2r}(\w) & := & \displaystyle{\sum_{k \in \cK} R_k \Phi^{M_r}(\w_{rk}) - C_r}  
  & \forall r \in \cR, \\
  g_{3k}(\w) & := & \displaystyle{\sum_{r \in \cR} \Phi^{M_r}(\w_{rk}) - 1}
  &  \forall k \in \cK, 
\end{array}
\end{equation*}
are the objective and constraint functions of~\eqref{eq:reform}. The AL method is iterative and generates a sequence of points $\{\w^i
\}$. The details of this method are given in Algorithm~\ref{alg:auglag}.
\begin{algorithm}[H]
  \caption{Augmented Lagrangian Method for Problem~\eqref{eq:reform}}
  \label{alg:auglag}
  1: (Initialization) Let $\mu_{\max} > 0$, $\beta > 0$ and $\lambda \in
  (0,1)$.  Choose initial multipliers $\mu_1^0, \mu_2^0
  \in [0,\mu_{\max}]^{R}$, $\mu_3^0 \in [0,\mu_{\max}]^{K}$
  and the penalty parameter $\rho_0 > 0$. Set $i \leftarrow 0$. \\
  2: (Stopping criteria) If $\w^{i}$ is a Karush Kuhn Tucker (KKT) point, then stop. \\
  3: (Subproblem) Find $\w^{i}$ as a solution of sub-problem (\ref{eq:subprobLagran})
 \begin{equation} 
  	\min_{\w} L_{\rho_i} (\w,{\mu}_1^i,{\mu}_2^i,{\mu}_3^i).
\label{prob:subprobLagran}
  \end{equation}
  4: (Update the Lagrange multipliers) Let
  \[
  \begin{array}{llll}
  \mu^{i+1}_{jr} & \leftarrow & \min\{\max \big\{ \mu^i_{jr} + \rho_i g_{jr} (\w^i), 0 \big\}, \mu_{max}\}
  & \forall r \in \cR, \quad j = 1,2, \\
  \mu^{i+1}_{3k} & \leftarrow & \min\{\max \big\{ \mu^i_{3k} + \rho_i g_{3k} (\w^i), 0 \big\}, \mu_{max}\}
  & \forall k \in \cK. \\
  \end{array}
  \] \\
  5: (Update the penalty parameter) Define 
  \[
  \begin{array}{llll}
    V^i_{jr} & = & \max \{ g_{jr}(\w^i), -\mu^i_{jr}/\rho_i \}
    & \forall r \in \cR, \quad j = 1,2, \\
    V^i_{3k} & = & \max \{ g_{3k}(\w^i), -\mu^i_{3k}/\rho_i \}
    & \forall k \in \cK, \\
  \end{array}
  \]
  If $i=0$ or
  \[
  \max \left\{ \max_{r \in \cR} V^i_{1r}, \max_{r \in \cR} V^i_{2r}, \max_{k \in \cK} V^i_{3k} \right\}
  \le \lambda \max 
  \left\{ \max_{r \in \cR} V^{i-1}_{1r}, \max_{r \in \cR} V^{i-1}_{2r}, \max_{k \in \cK} V^{i-1}_{3k} \right\},
  \]
  set $\rho_{i+1} = \rho_{i}$. Otherwise $\rho_{i+1} \leftarrow \beta \rho_i$. \\
  Let $i \leftarrow i + 1$ and return to Step~2.
\end{algorithm}

\subsection{Heuristic Algorithm for EE-Maximization}
\label{heuristic}
To take advantage of the specific F-RAN architecture, our proposed heuristic algorithm for energy-efficiency maximization incorporates a hybrid feature of cloud-centralized user association and F-AP distributed beamforming. This framework is similar to the one adopted in \cite{ourpaper,kaneko2019user} for sum-rate maximization, where it was shown to efficiently cope with the adverse effects of outdated CSIs. Namely, the proposed heuristic algorithm consists of two phases. Phase I is carried out at the cloud BBUs for resolving the optimal user clustering and set of active F-APs based on the global but imperfect user CSI. In other words, the goal of Phase I is to solve for variable $\cR_A$. Then, based on the solution from Phase I, Phase II will optimize over the beamforming vectors $\w_{rk}$, $r \in \cR_A, k \in \cK_r$, at each F-AP, based on local but perfect CSI knowledge. Note that, unlike the initial problem (\ref{eq:reform}) which solves the global beamforming vector $\w$ then infers $\cR_A$, the decoupled resolution of these variables is a design choice enabling to take advantage of the centralized/distributed F-RAN architecture for reducing the required computational complexity.

A specific feature of our proposed heuristic algorithm is that Phase I is optimized periodically, namely, every $T$ frames, using outaded CSI in a centralized manner, while beamforming is optimized locally at every scheduling frame with perfect CSI. This allows us not only to reduce transport delays, but also to save a large amount of power  for transmitting CSI and control  information from F-APs to the cloud BBUs. 

The details of the proposed heuristic algorithm are described as follows.
  
\emph{\textbf{For Frame 1 among T scheduling frames: }}
As mentioned, given the difficulty of optimizing the energy efficiency objective function jointly over beamforming vectors and active F-AP set, problem (\ref{eq:EEproblem}) is decomposed and addressed through two phases. And in this first frame, both Phase I and Phase II will be performed, where Phase I includes two steps for resolving the optimized set of active F-APs while Phase II resolves for optimized beamforming vectors.\\[1.5mm]
\textbf{Phase I - Step 1: } \underline{Reformulation as a sum-rate maximization problem for initial user association}\\[1.5mm]
In this first step, in order to make problem (\ref{eq:EEproblem}) tractable, we reformulate it into an equivalent sum-rate maximization problem under simplifying assumptions, and solve for $\w_{rk}$. To do this, we assume that all F-APs are activated, i.e., the set of active F-APs is fixed to $\cR_A=\cR$, and the transmit power $P_r^w$ in (\ref{eq4}) is assumed to be equal to its maximum budget $P_r^{\max}$ in (\ref{eq:PrConstraint}). This sets the denominator in (\ref{eq:EEproblem}) to its maximum value, which is a constant. By doing so, problem (\ref{eq:EEproblem}) is recast as the following sum-rate maximization problem,

\begin{IEEEeqnarray}{lrl}
\label{eq:SR} \IEEEyessubnumber*
&\max_{{\w}_{rk}} \quad \tau  = & \quad \sum_{k \in \cK}R_k,
\nonumber \IEEEyesnumber\\
\text{s.t.,} \quad &	\sum_{k \in \cK_r} ||\w_{rk}||_2^2 \   \leq  \ &  P_r^{\max}, \quad \hfill \forall r \in \cR, \label{eq:PrConstraintSR} \IEEEyessubnumber \\
& \sum_{k \in \cK_r}R_k \ \leq \  &  C_r, \hfill  \forall r \in \cR, \label{eq:CrConstraintSR} \IEEEyessubnumber \\
&\sum_{r \in \cR_A}\Big\|||\w_{rk}||_2^2\Big\|_0 \ \leq \ &  1, \hfill \forall k \in \cK.
\label{eq:FogConstraintSR}\IEEEyessubnumber
\end{IEEEeqnarray}
Note that, due to the fronthaul delays, the data rate is calculated by $R_k = \log(1 + \tilde{\gamma}_k)$, where $\tilde{\gamma}_k$ denotes the SINR of user $k$ which is a function of the outdated CSI $\tilde{\h}_k$ in (\ref{eq:outdateCSI}).

To tackle problem (\ref{eq:SR}), Algorithm \ref{al:PreSchedule} below is applied, whereby the following relaxation technique is introduced in order to approximate constraint (\ref{eq:FogConstraintSR}) as 
\begin{equation}
\sum_{r \in \cR}\Big\|||\w_{rk}||_2^2\Big\|_0 \approx \sum_{r \in \cR}\beta_{rk}||\w_{rk}||_2^2 \leq 1.
\label{eq:approx2}
\end{equation}
Here,
\begin{equation}
\beta_{rk} = \frac{1}{||\hat{\w}_{rk}||_2^2 + \delta},
\label{eq:appx2}
\end{equation}
with $\hat{\w}_{rk}$ the value of $\w_{rk}$ at the algorithm's previous iteration. Parameter $\delta > 0$ is a regularization factor that is tuned by (\ref{eq:tuning}) such that it diminishes over iterations and $\beta_{rk}||\w_{rk}||_2^2$ approaches $\Big\|||\w_{rk}||_2^2\Big\|_0$ at convergence: 
\begin{equation}
\label{eq:tuning}
\delta_i = \delta_0\lambda^i.
\vspace{-2mm}
\end{equation}
In (\ref{eq:tuning}),  $\delta_0$ is an appropriate initial value of $\delta$ and $\lambda \in (0, 1)$. Actually, the approximation of constraint (\ref{eq:FogConstraintSR}) by (\ref{eq:approx2}) is similar to the approximation by (\ref{eq:appx}), but with the difference that, by using $\beta_{rk}$ calculated with the previous value of $\w_{rk}$ in (\ref{eq:appx2}), constraint (\ref{eq:FogConstraintSR}) becomes convex. 
Then, problem (\ref{eq:SR}) becomes a convex optimization problem that can be solved by standard methods, for instance the interior point method. The initial user clustering is hence obtained through the solution of this convex optimization problem by observing that $k \in \cK_r$ if and only if $\w_{rk} \neq \bm{0}$.

\begin{algorithm}[h!]
\caption{Initial User Association at BBUs}
	\begin{algorithmic}[1]
		\State \textbf{Initialize} $\{\hat{\w}_{rk}, \delta\}$
		\Repeat
			\State Update $\beta_{rk}$ 
			\State Update $\delta = \delta \cdot \lambda$
			\State For fixed $\beta_{rk}$, cast (\ref{eq:SR}) as an SOCP and solve by the interior point method.
		\Until{$||\hat{\w}_{rk} - \w_{rk}||_2^2 < \epsilon$}
	\end{algorithmic}
\label{al:PreSchedule} 
\end{algorithm}

\noindent \textbf{Phase I - Step 2: }\underline{Energy-efficient greedy F-AP activation for solving $\cR_A$} \\[1.5mm]
We define the global energy efficiency metric $\eta^{Glo}$, and the local energy efficiency metric $\eta_r^{Loc}$ of each F-AP $r$ as follows, based on the initial solutions $\w_{rk}$ obtained in step 1,
\begin{equation}
\label{eq:GloEE}
	\eta^{Glo} = \frac{B\tau(\w_{rk}, \cR_A)}{P(\w_{rk}, \cR_A)} =\frac{B\sum_{k \in \cK}R_k}{P(\w_{rk}, \cR_A)},
\end{equation}
\begin{equation}
\label{eq:LocalEE}
	\eta_r^{Loc} = \frac{B\tau_r(\w_{rk}, \cR_A)}{P_r^{\text{on}}(\w_{rk}, \cR_A)} =\frac{B\sum_{k \in \cK_r}R_k}{P_r^{\text{on}}(\w_{rk}, \cR_A)},
\end{equation}
with $P$ and $P_r^{\text{on}}$ defined in (\ref{eq:power}), (\ref{eq:PF}) respectively. The global energy efficiency metric englobes all F-APs' contributions, while the local energy efficiency metric only accounts for F-AP $r$. 

F-APs are then listed in the descending order of $\eta_r^{Loc}$, and the lowest F-AP is removed from the set of active F-APs $\cR_A$ if  $\eta^{Glo}$ is increased. This greedy process is iterated, until $\eta^{Glo}$ stops increasing (lines 11 to 19 in Algorithm \ref{al:GeneralAlgorithm}). At the end, the energy-efficient solutions for user association $\cK_r, \forall r \in \cR$  and the set of active F-APs $\cR_A$ are obtained.\\[2mm] 
\textbf{Phase II: } \ul{Signal to Leakage-plus-Noise Ratio (SLNR) maximization for solving $\w_{rk}$} \\[1.5mm]
In this phase, based on their associated users given by the previous phase, each active F-AP optimizes beamforming, using local but perfect knowledge of CSI. Given this local context, we opt for maximizing the SLNR metric at each user served by active F-APs for optimizing the actual beamforming vectors $\w_{rk}$. The SLNR of user $k$ associated to F-AP $r$ towards other users is given by
\begin{equation}
	\zeta_{rk} = \frac{|\h_{rk}^H\w_{rk}|^2}{\sum\limits_{\substack{k' \in \cK \\ k' \neq k}}|\h_{rk'}^H\w_{rk}|^2 + \sigma_n^2}.
\label{eq:SLNR}
\end{equation}
Then, under the assumption that the power allocation among users associated to the same F-AP is equal, the SLNR maximization problem at each active F-AP is expressed as 
\begin{equation}
\max_{\w_{rk}} \zeta_{rk} \quad \text{such that} \quad ||\w_{rk}||_2^2 \leq \frac{P_r}{K_r}.
\label{eq:solution}
\end{equation}
This problem can be solved in closed-form, where the optimal beamforming vectors as given by \cite{sadek_active_2007}, 
\begin{equation}
\w_{rk}^{opt}=\sqrt{\frac{P_r}{K_r}} \max eig\Bigg[\left(\sum\limits_{k'\neq k}\h_{rk'}\h_{rk'}^H+ \frac{K_r{\sigma_n^2}}{P_r}\I\right)^{-1}\h_{rk}\h_{rk}^H\Bigg],
\end{equation}
where $\max eig(A)$ is the eigenvector of the largest eigenvalue of matrix $A$, and can be efficiently computed by the power iteration method \cite{golub2012matrix}.

Finally, for the first frame, the total consumed power at each F-AP is given by $P_r$ in (\ref{eq:power}). 

\emph{\textbf{For Frame 2 to T: }}
For these intermediate scheduling frames, the user association and set of active F-APs will remain fixed to that of the first frame, so that only beamforming optimization, i.e., Phase II, is performed individually at each active F-AP, based on local and perfect CSIs available for each frame. Therefore, the energy consumption of each F-AP solely includes circuit power $P_r^c$ and wireless downlink transmit power $P_r^w$. 

The overall power consumption for our proposed algorithm, averaged over the $T$ scheduling frames is finally given by
\begin{equation}
P_{prop} = \frac{1}{T} \bigg[\Big(\sum_{r \in \cR_A}(P_r^{f,u} + P_r^c + P_r^{f,d} + P_r^{w}\Big) + \sum_{t = 1}^{T - 1}\sum_{r \in \cR_A}(P_r^c + P_r^{w}) + \sum_{t = 1}^{T}\sum_{r\notin \cR_A}(P_r^{f,u} + P_r^c)\bigg].  \label{eq:PF2}
\end{equation}

Details of the proposed algorithm are described in Algorithm \ref{al:GeneralAlgorithm}.
\vspace{-2mm}
\begin{algorithm}
\caption{Proposed Heuristic EE Algorithm for solving (\ref{eq:EEproblem})}
\label{al:GeneralAlgorithm}
	\begin{algorithmic}[1]
		\State Time slot $t \gets 0$
		\While{$t < MAX\_TIME\_SLOT$}
			\If{$t$ mod $T$ = 0}
				\State $\cR_A \gets \cR$
				\State $\{\cK_1, \ldots, \cK_R\} \gets$  Solve (\ref{eq:SR})
				\State Calculate $\eta^{Glo}$ by (\ref{eq:GloEE})
				\For{each $r \in \cR_A$}
					\State Calculate $\eta_r^{Loc}$ by (\ref{eq:LocalEE})
				\EndFor
				\State $\cR_A' \gets \cR_A$
				\While{$true$}
					\State $r_{wst} \gets \arg\min_{r \in \cR_A'}\{\eta_r^{Loc}\}$
					\State Calculate $\eta^{Glo'}$ by (\ref{eq:GloEE}) with $\{\cR_A \backslash r_{wst}\}$
					\If{$\eta^{Glo'} \geq \eta^{Glo}$} 
						\State $\cR_A' \gets \cR'_A \backslash r_{wst}$	
					\Else
						\State $\cR_A \gets \cR'_A$, \textbf{break}
					\EndIf
				\EndWhile
			\EndIf
			\State	Perform local beamforming optimization for every F-AP $r \in \cR_A$
			\State $t \gets t + 1$
		\EndWhile
	\end{algorithmic}
\end{algorithm}

%% file: results.tex
\subsection{Simulation settings}
\label{simulationSystem}
We evaluate the proposed and reference algorithms in two types of heterogeneous F-RAN networks: firstly, a small-scale network composed of 1 macro and 3 pico F-APs with 5 and 10 users, and secondly, a larger network composed of 3 macro and 9 pico F-APs with 15 to 90 users. Users are uniformly distributed over the whole network area. The number of antennas of the macro and the pico F-APs are equal to  4 and 2, respectively. Each user is equipped with one antenna.

The wireless channels are assumed to undergo small-scale Rayleigh fading and log-normal shadowing. Simulation parameters generally follow those of \cite{dai_backhaul-aware_2015}. Namely, the maximum transmit power budgets of macro and pico F-APs are set to 43 and 30 dBm respectively, and the fronthaul rate limitation $C_r$ is fixed to 690 Mbps  for macro F-APs and 107 Mbps for pico F-APs. The bandwidth $B$ is set to 10 MHz and the noise power spectral density $\sigma_n^2$ to $-169$ dBm/Hz. Regarding the power consumption model, parameter values follow those of \cite{na_energy_2018} and are given in Table~\ref{tb:simulatepara}.

\begin{table}[tbh!]
\caption{Simulation parameter values for the power consumption model \cite{na_energy_2018}}
\centering
\begin{tabularx}{0.5\textwidth}{c | X }
\hline \hline
\textbf{Parameter} & \textbf{Description} \\ \hline
$P_{\textrm{fix},r}$ & 0.825 W\\ \hline
$\beta$ & 4/3\\ \hline
$b_{IQ}$ & 20\\ \hline
$f_{\textrm{pre}}$ & 1.5 MHz\\ \hline
$P_{\textrm{td},r}$ & 0.25 W/Gbps \\ \hline
$\beta_r$ & 0.4\\ \hline
$\rho_dN_0$ & 1 \\ \hline
$P_{\textrm{ic},r}$ & 0.2 \\
\hline \hline
\end{tabularx}
\label{tb:simulatepara}
\end{table}

All simulations are run with Matlab R2018a on an Intel Core i7-7700 3.6 Ghz processor.

In the proposed AL-based EE algorithm, denoted \emph{Prop. AL} algorithm, we set $\mu_{\max}$ to 1, $\lambda$ to 0.25, $\beta$ to 10, the initial value of penalty parameter $\rho_0$ to 10 and the maximum number of iterations to 50. The value of $\delta$ in both $\ell_0$-norm approximations (\ref{eq:appx}), (\ref{eq:appx2}) is chosen as 0.1. The scheduling period of the proposed heuristic EE scheme, denoted as \emph{Prop. Heur.} algorithm, is set to $T = 10$ as in \cite{ourpaper,kaneko2019user}. 

To evaluate the energy efficiency performance, both proposed algorithms are compared to the reference algorithm designed in \cite{na_energy_2018} for providing high energy-efficiency in F-RAN, referred as \emph{Ref. EE} algorithm and described in subsection \ref{subsec:refAlg} for convenience. However, due to its aforementioned high computational complexity, \emph{Prop. AL} algorithm will be only considered in the small-scale network, whereas both \emph{Prop. Heur.} and \emph{Ref. EE} will be evaluated in the larger network. 

Moreover, since sum-rate is a key performance metric which should not be sacrificed even when aiming at high global energy efficiency, the proposed heuristic algorithm was also compared to the sum-rate maximization algorithm of \cite{ourpaper,kaneko2019user} designed for F-RAN, which essentially performs user association and beamforming as in Step 1 of section V.B. This algorithm will be referred to as \emph{Ref. SR} algorithm.

To assess the effects of various CSI error levels, we evaluate the proposed heuristic algorithm for different CSI error variances $\sigma_e^2 = \{0, 0.01, 0.1, 1\}$, where $\sigma_e^2 = 0$ represents the perfect CSI case. 

\subsection{Reference algorithms}
\label{subsec:refAlg}
\subsection*{\textbf{Reference energy-efficiency maximizing algorithm for F-RAN (Ref. EE)}}
This method, proposed in \cite{na_energy_2018}, leverages edge processing for enabling a dynamic F-AP selection to enhance system energy efficiency. 
Each F-AP autonomously decides to turn off or not, based on its local but perfect CSI. At the beginning, each user chooses its closest F-AP based on its received signal power strength from F-APs, then if the number of associated users to a given F-AP surpasses its capacity, the worst users in terms of energy efficiency will be dropped. After that, if the number of associated users is lower than a fixed threshold, the F-AP will be turned off. In \cite{na_energy_2018}, beamforming optimization is performed at each F-AP towards its local users based on zero-forcing, however, SLNR-based optimization will be performed here for fair comparison with our proposed scheme\footnote{Preliminary simulations showed that both beamforming methods gave similar results for \emph{Ref. EE} algorithm.}.

The details of this method are given in Algorithm \ref{al:refCCNC}. The value of $\theta$ is set to 1 in our simulation, i.e., an F-AP is turned off if it has less than 1 associated user. 

\begin{algorithm}
\caption{\emph{Ref. EE Algorithm \cite{na_energy_2018}}}
\label{al:refCCNC}
	\begin{algorithmic}[1]
		\State Initialization: $\cK_r = \cK$ for $\forall r$, $\cR_A = \cR$
		\For{each $r \in \cR_A$}
			\While{$|\cK_r|$ is over capacity of $r$}
				\State $\tilde{k} = \arg\max_{k\in \cK_r}d_{kr}$ \Comment{$d_{kr}$: distance between user $k$ and $r$}
				\State $\cK_r = \cK_r \backslash \{\tilde{k}\}$
			\EndWhile
			\If{$|\cK_r| < \theta$} \Comment{$\theta$: the minimum served users by F-AP}
				\State $\cR_A = \cR_A \backslash \{r\}$
			\EndIf
		\EndFor
		\State Calculate global energy efficiency with updated $\cR_A$ and $\cK_r$, $\forall r \in \cR_A.$
	\end{algorithmic}
\end{algorithm}

\subsection*{\textbf{Reference sum-rate maximizing algorithm for F-RAN (Ref. SR)}}
In \cite{ourpaper,kaneko2019user}, the issue of user scheduling and beamforming in F-RANs was also considered, but regarding weighted sum-rate maximization of the whole system. This algorithm is also composed of  two phases: Phase I solves for the set of active F-APs using global, but outdated CSI and Phase II solves for the beamforming vectors at each active F-APs using local, but perfect CSI. However, in Phase I which is performed in frame 1 among $T$, only steps 1 and 3 are implemented by the sum-rate reference algorithm. Thus, the simulation results will show the efficiency of our F-APs' activation strategy, which is performed in Step 2 of Phase I of our proposed heuristic algorithm. 
\subsection{Simulation results}
\label{results}
\subsection*{\textbf{Small-scale network}}
\begin{figure*}[tb!]
\centering
\begin{subfigure}{.33\textwidth}
\includegraphics[scale=0.38]{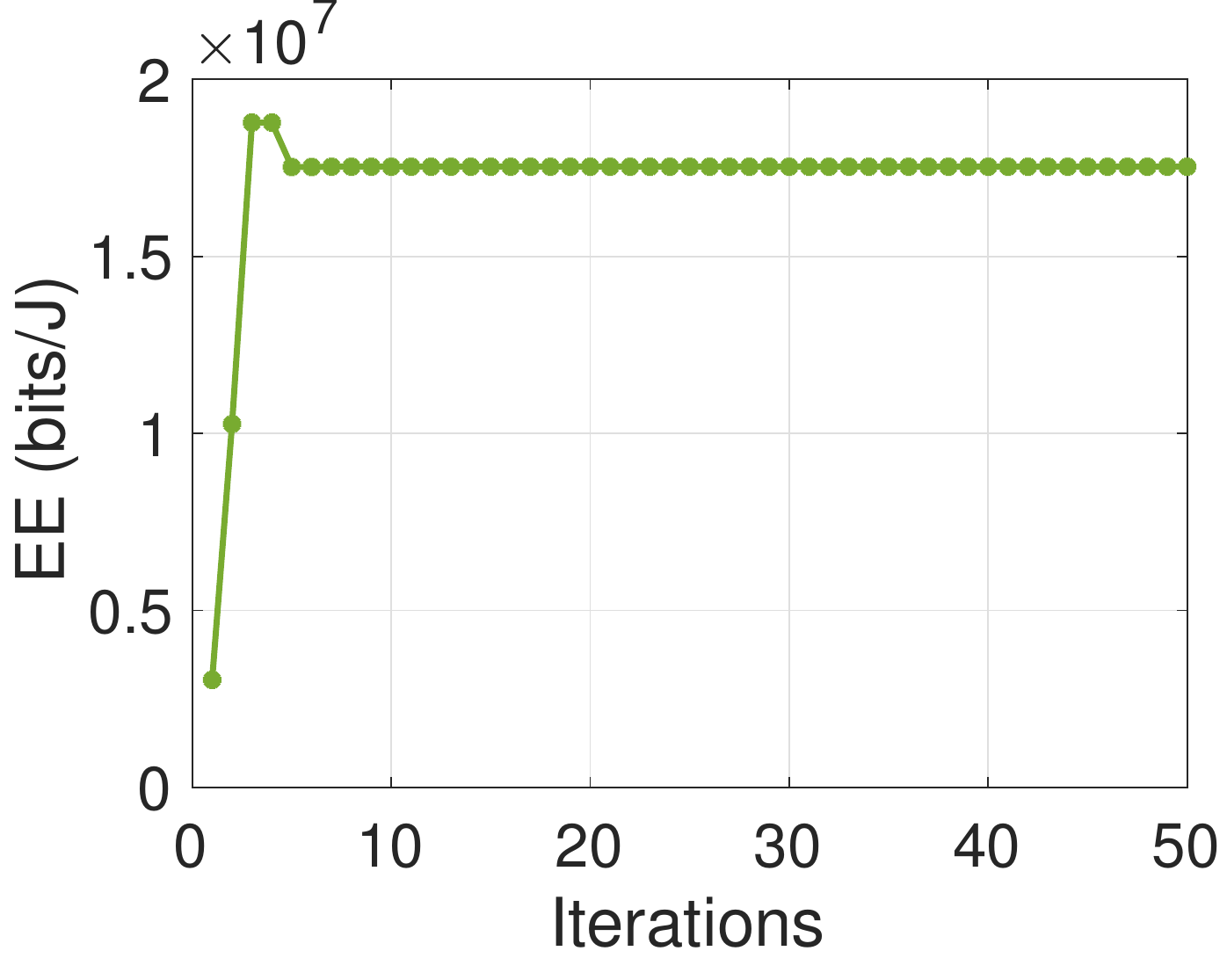}%
\caption{}
\label{fig:ALconv1}
\end{subfigure}%
\begin{subfigure}{.33\textwidth}
\includegraphics[scale=0.38]{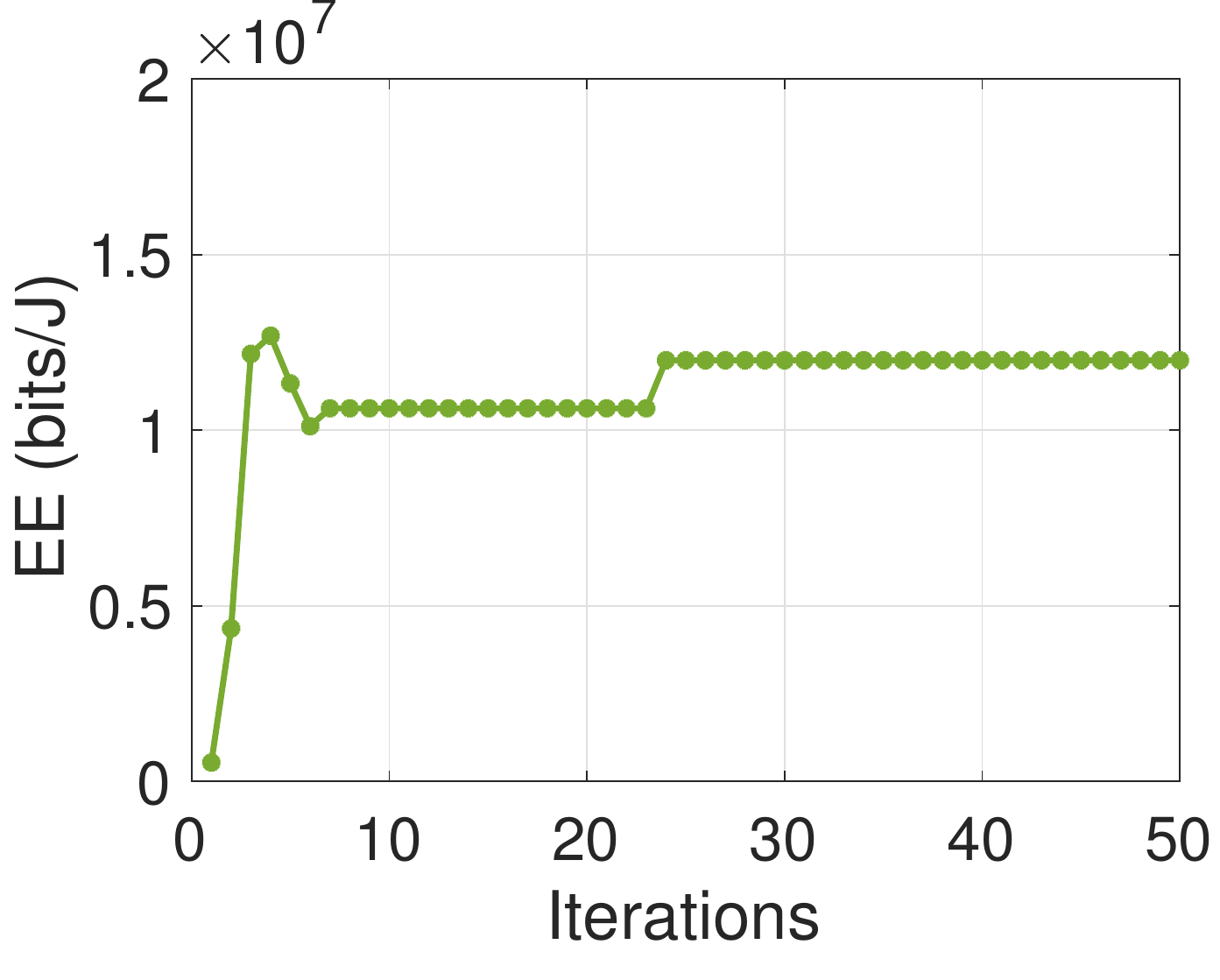}%
\caption{}
\label{fig:ALconv2}
\end{subfigure}%
\begin{subfigure}{.33\textwidth}
\includegraphics[scale=0.38]{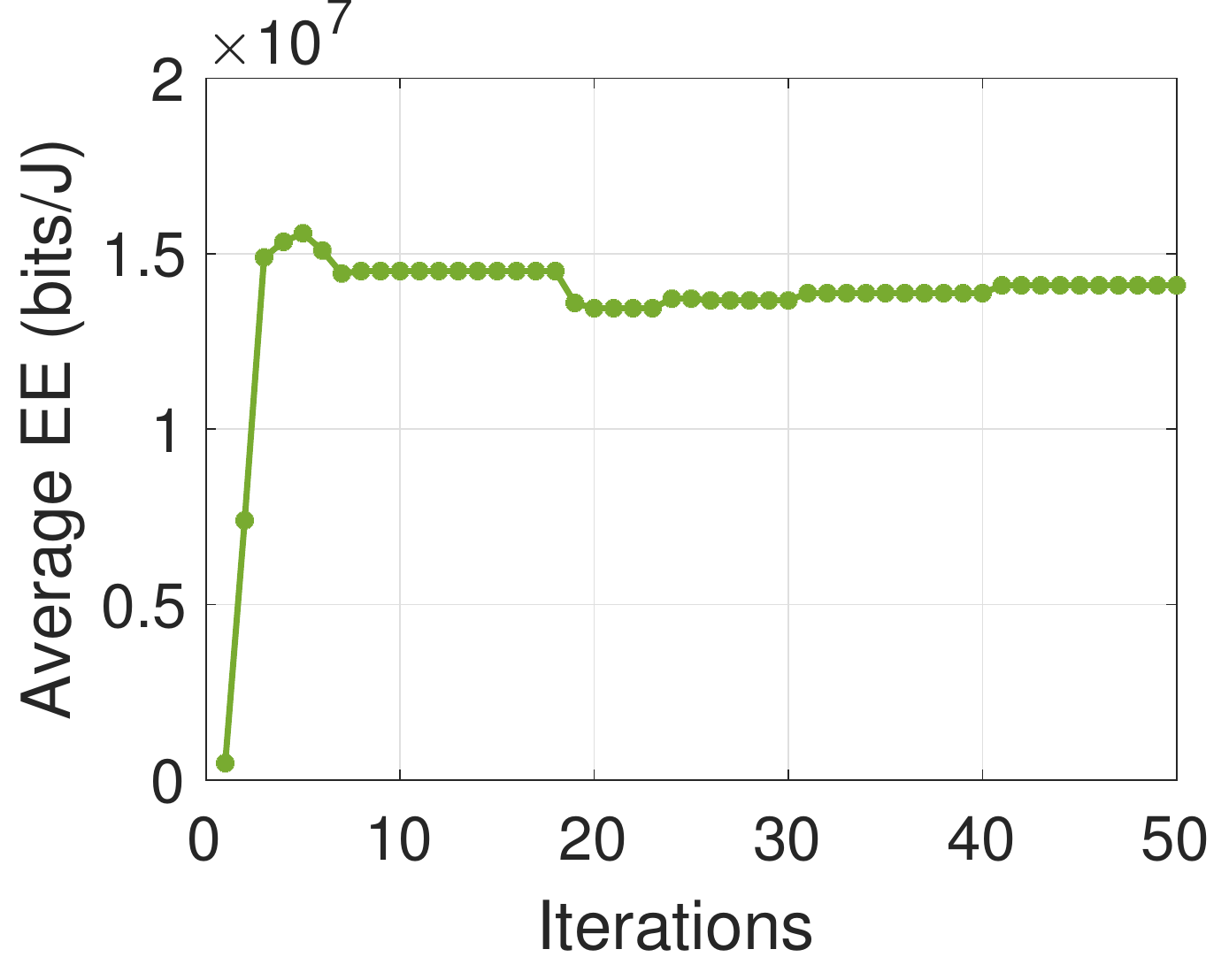}%
\caption{}
\label{fig:ALconv3}
\end{subfigure}%
\caption{Convergence behavior of the proposed AL-based algorithm}
\label{fig:ALconvergence}
\end{figure*}

As mentioned above, due to its high computational complexity, the proposed AL-based algorithm, \emph{Prop. AL}, is only considered in a small-scale network. This algorithm has the advantage of serving as a performance upper-bound to assess the efficacy of the proposed heuristic algorithm. First, we show the convergence behavior of \emph{Prop. AL} in Fig. \ref{fig:ALconvergence}, where Figs. \ref{fig:ALconv1} and \ref{fig:ALconv2} show the behavior of  \emph{Prop. AL} at two different channel instances as examples, while Fig. \ref{fig:ALconv3} represents its convergence behavior after averaging over multiple channel realizations. We can observe that, in all cases,  \emph{Prop. AL} converges well after around 25 iterations.

Next, we compare the proposed and reference algorithms' performances, assuming perfect CSI. 
Obviously, Table \ref{tb:EEresults} shows that both proposed algorithms outperform the reference scheme, achieving more than twice and 1.6 times higher EE performance for \emph{Prop. AL} and \emph{Prop. Heur.}, respectively. In addition, the proposed heuristic algorithm provides an appreciable energy efficiency level, namely 75\% of  that achieved by the proposed AL-based method.

\begin{table}
\centering
\caption{Average energy efficiency [Mbits/J] for proposed and reference algorithms in the small-scale network}
	\setlength{\tabcolsep}{0.5em}	
	{\renewcommand{\arraystretch}{1.1}%
	\begin{tabular*}{0.5\columnwidth}{r|ccc}
		\hline \hline
		&\emph{Prop. AL} & \emph{Prop. Heur.}& \emph{Ref. EE} \\ \hline
		5 users  & 21.55 & 16.51 & 10.24\\
		10 users & 23.17 & 16.86 & 10.94 \\ 
		\hline \hline
	\end{tabular*}}
\label{tb:EEresults}
\end{table}

\begin{table}
\centering
\caption{Average running time [second] per frame for proposed and reference algorithms in the small-scale network}
	\setlength{\tabcolsep}{0.5em}	
	{\renewcommand{\arraystretch}{1.1}%
	\begin{tabular*}{0.5\columnwidth}{r|ccc}
		\hline \hline
		&\emph{Prop. AL} & \emph{Prop. Heur.}& \emph{Ref. EE} \\ \hline
		5 users  & 202.2 & 0.36 & 0.17\\
		10 users & 415.3 & 0.53 & 0.33 \\ 
		\hline \hline
	\end{tabular*}}
\label{tb:runningtime}
\end{table}

\begin{figure*}[t]
\centering
\begin{subfigure}{.49\textwidth}
\includegraphics[scale=0.49]{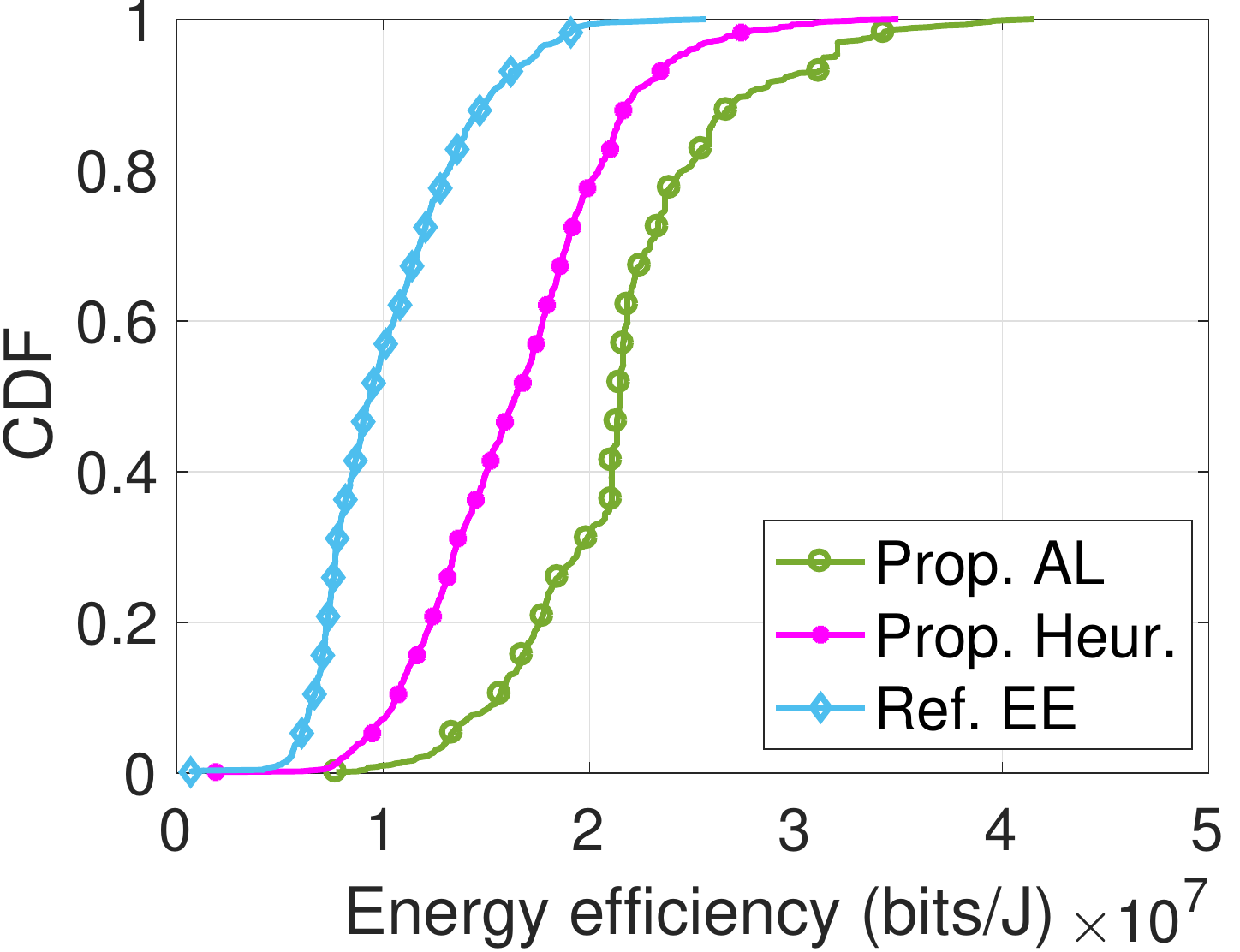}%
\label{fig:AL1}
\caption{5 users}
\end{subfigure}%
\begin{subfigure}{.49\textwidth}
\includegraphics[scale=0.49]{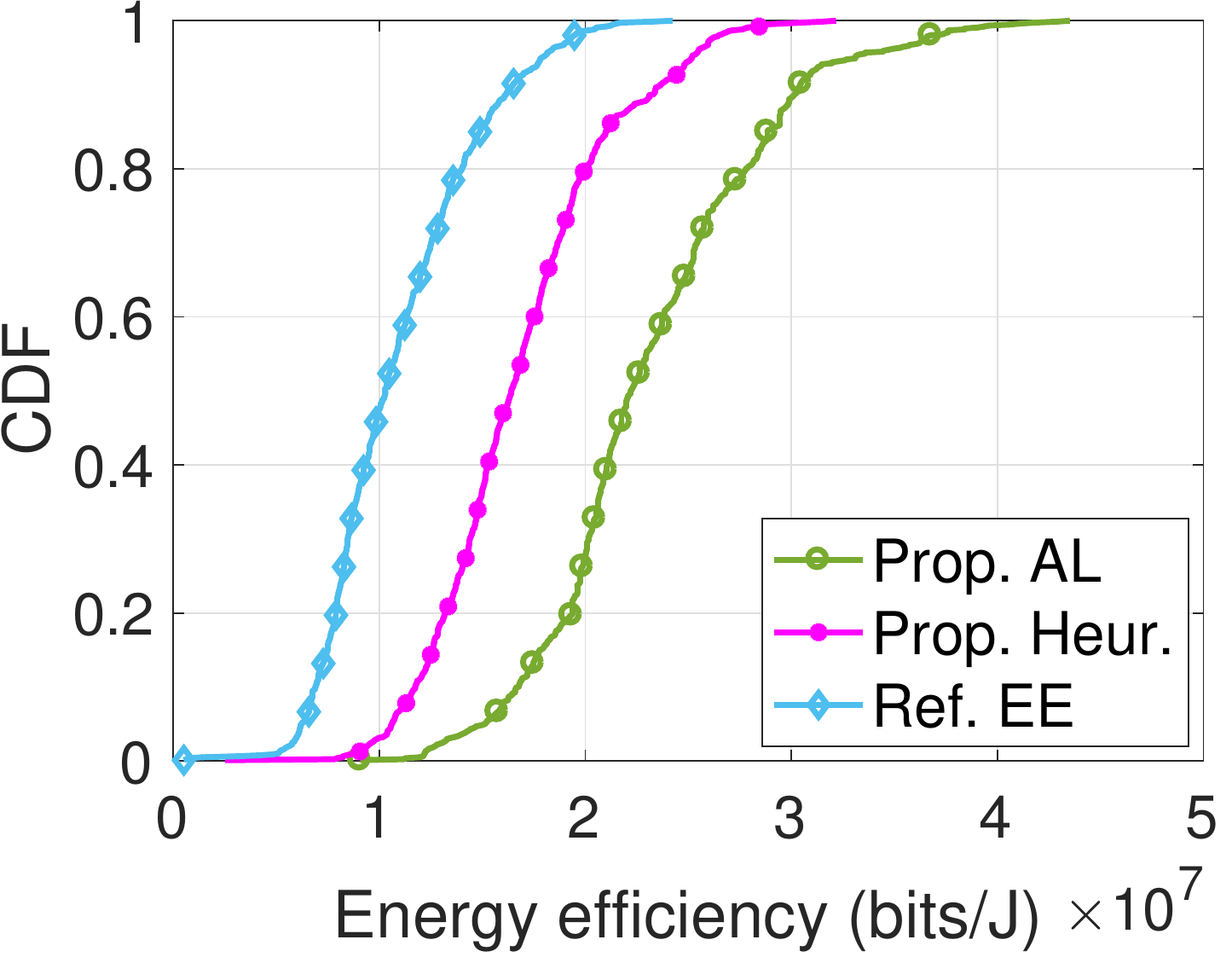}%
\label{fig:EERef1}
\caption{10 users}
\end{subfigure}%
\caption{CDF of energy efficiency given by proposed and reference algorithms in small-scale networks with perfect CSI}
\label{fig:CDFEEsmall}
\end{figure*}

Moreover, Fig. \ref{fig:CDFEEsmall} shows the cumulative distribution function (CDF) of energy efficiency given by proposed and reference algorithms, which again shows the clear gains achieved by both proposed algorithms, compared to the reference algorithm, namely in the case of 5 users, 92\% and 54\% gains at 80-th percentile for \emph{Prop. AL} and \emph{Prop. Heur.}, respectively, and 100\% and 43\% gains at the same percentile for \emph{Prop. AL} and \emph{Prop. Heur.} in the case of 10 users.  

In addition, Table \ref{tb:runningtime} summarizes the average running times required per frame, for all algorithms, which clearly shows the high computational complexity of \emph{Prop. AL}. Namely, from this table, we can see that in the case of 5 users, \emph{Prop. AL} takes 560 and 1000 times longer running times compared to \emph{Prop. Heur.} and \emph{Ref. EE} respectively, and this goes up to 790 and 1200 times longer running times in the case of 10 users, respectively. Hence, this confirms that \emph{Prop. AL} is intractable for large-scale networks.

\begin{figure*}
\centering
\begin{subfigure}{.33\textwidth}
\includegraphics[scale=0.38]{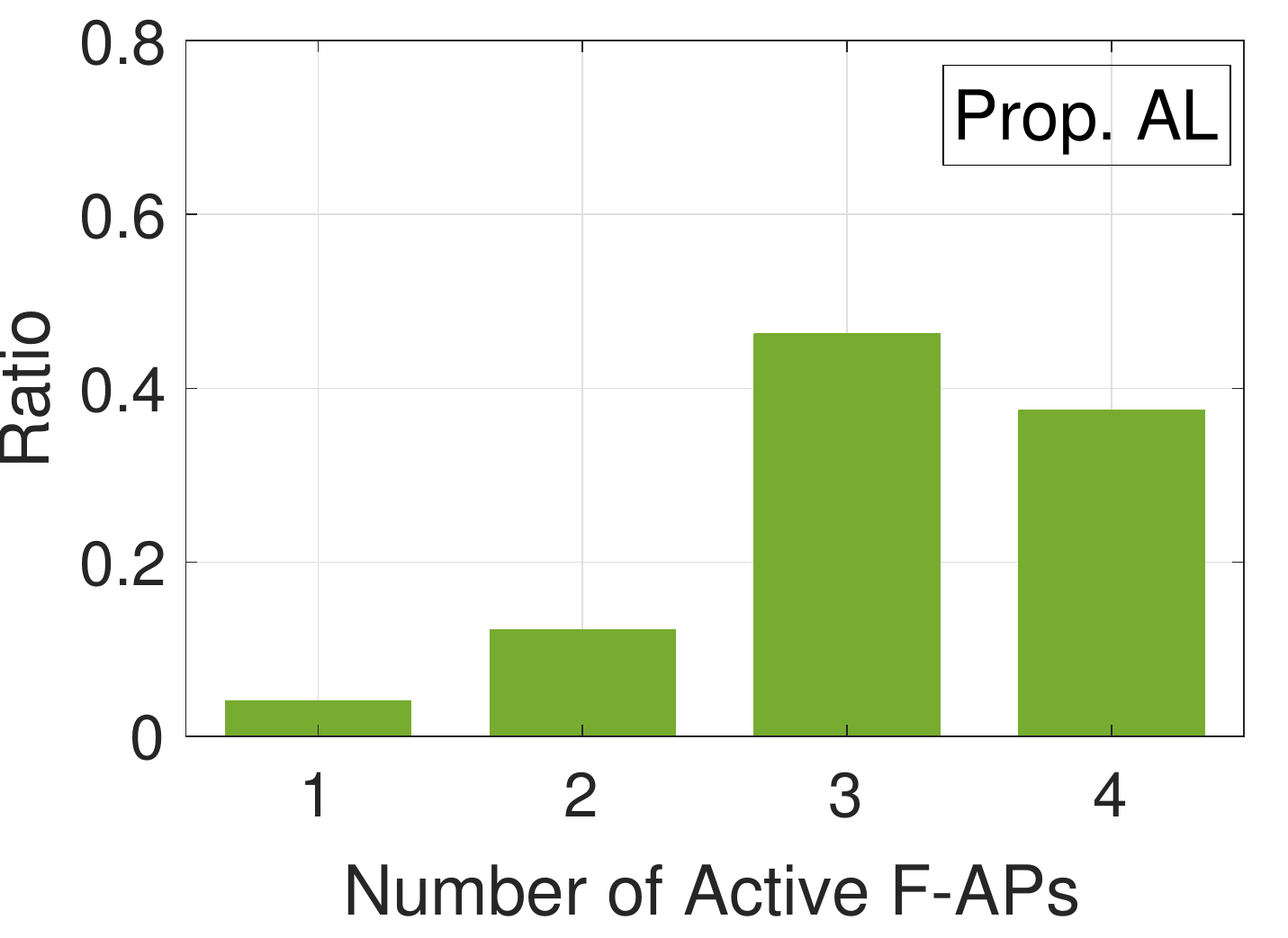}%
\label{fig:AL1}
\end{subfigure}%
\begin{subfigure}{.33\textwidth}
\includegraphics[scale=0.38]{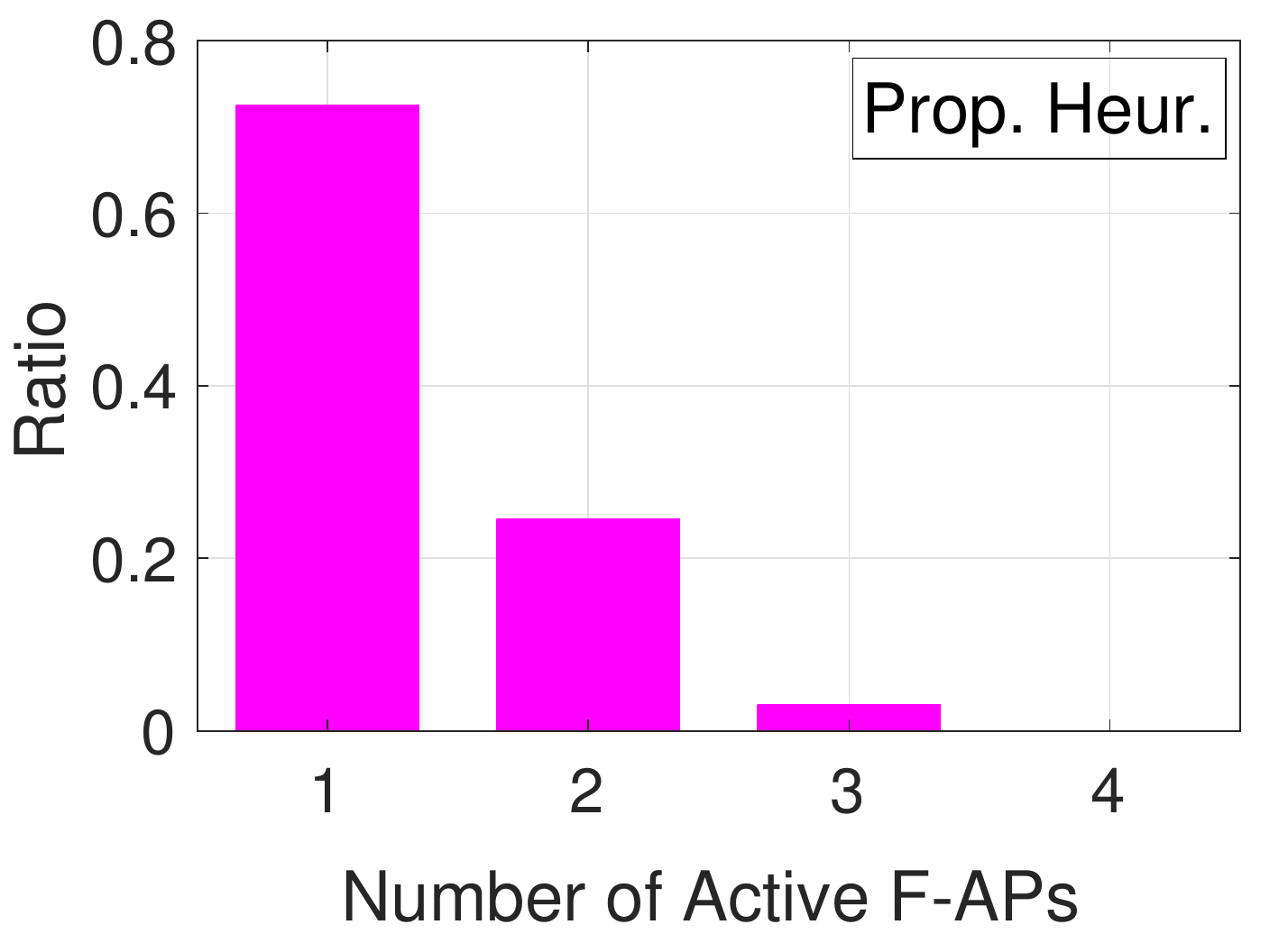}%
\label{fig:Heu1}
\end{subfigure}%
\begin{subfigure}{.33\textwidth}
\includegraphics[scale=0.38]{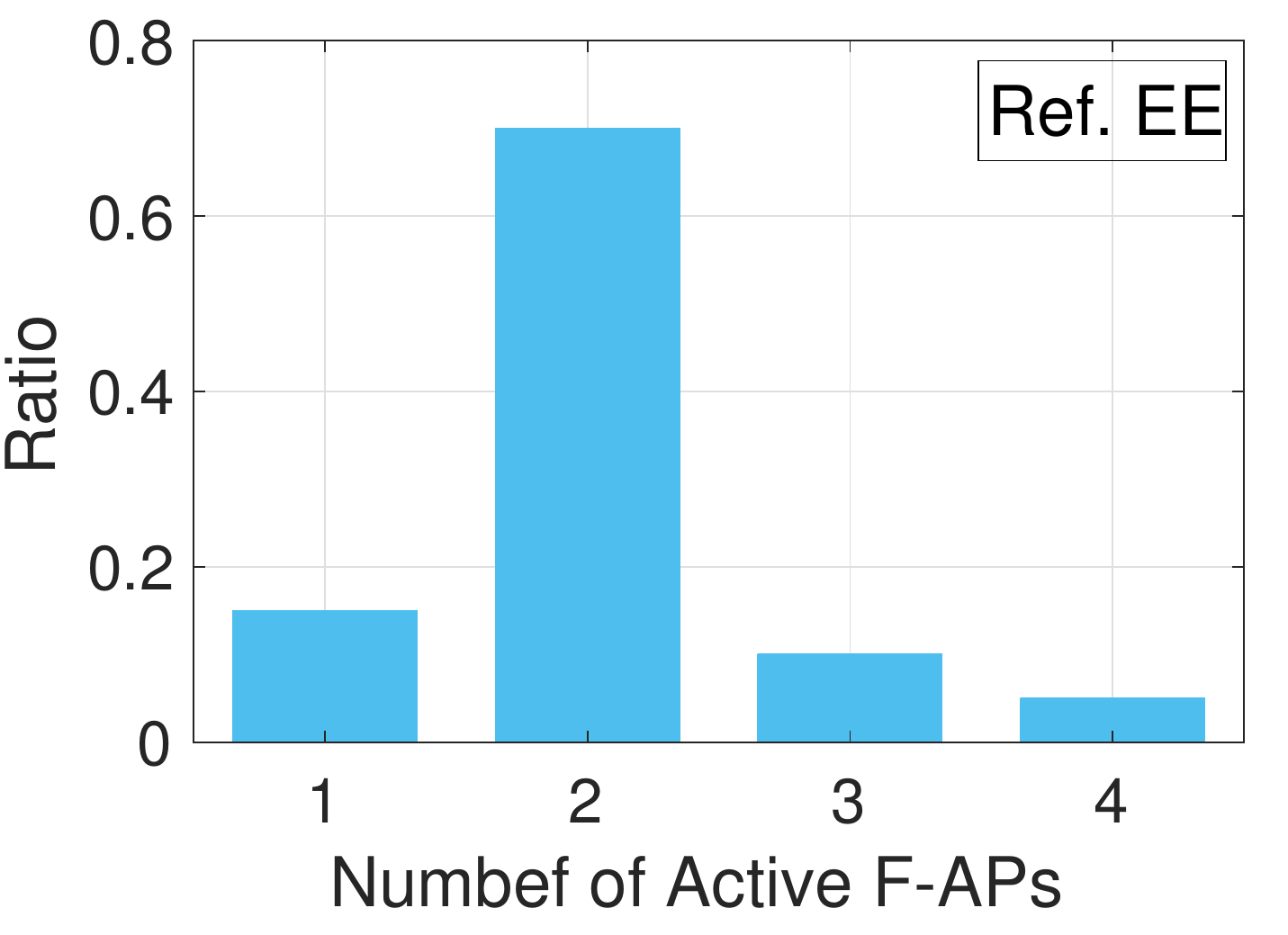}%
\label{fig:EERef1}
\end{subfigure}%
\caption{Distribution of the number of active F-APs, in a small-scale network, $K=10$ users}
\label{fig:nbRRH1}
\end{figure*}

Finally, in Fig. \ref{fig:nbRRH1}, we show the distribution of the number of active F-APs given by the three algorithms in the case of 10 users. It can be observed that the number of active F-APs in the heuristic and reference schemes is notably reduced compared to that of the optimization method. This is due to the fact that \emph{Prop. AL} solves first for the overall beamforming vector from all F-APs to users, then only F-APs whose beamforming vectors towards all users are zero ($\w_{rk} = \mathbf{0}$) are turned off. By contrast, both \emph{Prop. Heur.} and \emph{Ref. EE} turn off the low-energy efficient F-APs prior to solving for the actual  beamformers. 
 However, our F-APs activation strategy appears to be much more efficient than in the reference one, since we guarantee a higher energy efficiency performance despite a lower number of active F-APs. In addition, despite deactivating a larger portion of F-APs, \emph{Prop. Heur.} still provides higher user rate fairness as compared to \emph{Ref. EE}, as will be shown in the case of larger networks in the sequel. 

\subsection*{\textbf{Larger networks}}
In this case, we first compare our proposed heuristic algorithm with the reference method in terms of energy efficiency, under various CSI imperfectness levels at the cloud. 

%

\begin{figure}
\centering
\includegraphics[scale=0.45]{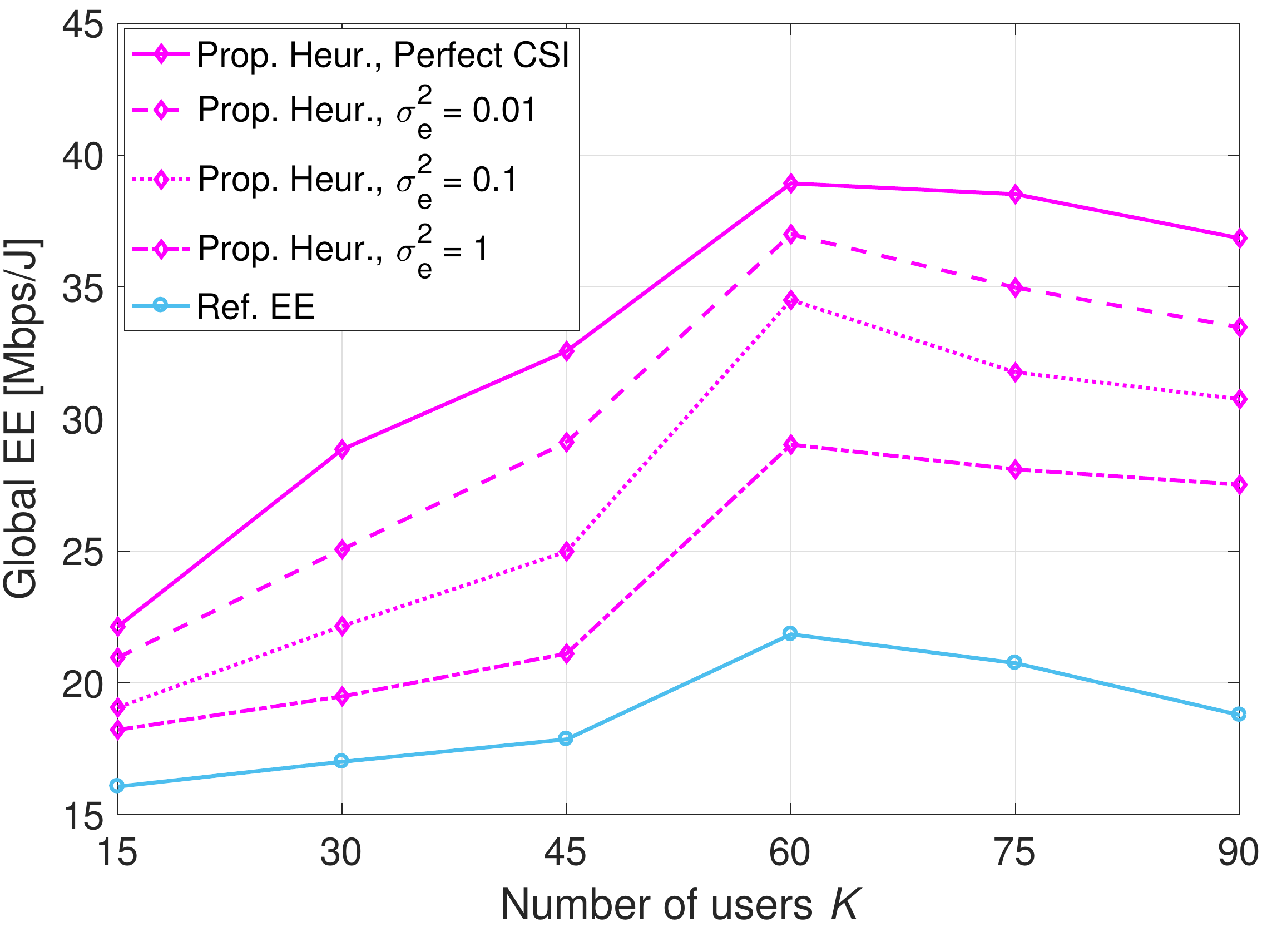}
\caption{Energy efficiency performance of the reference and proposed heuristic algorithms against number of users under various global CSI imperfectness levels}
\label{fig:EEall}
\end{figure}

Since the reference scheme carries out both user association and beamforming tasks at individual F-APs, its performance is not impacted by the CSI errors at the cloud. By contrast, in our proposed algorithm, user association is implemented at cloud BBUs, thereby reducing the achievable energy efficiency performance as the CSI error level increases. This can be observed clearly in Fig. \ref{fig:EEall} showing the four curves given by \emph{Prop. Heur.} under the four CSI error levels, but only one curve for \emph{Ref. EE} scheme. We can also note that, although the performance of \emph{Prop. Heur.} algorithm reduces as the CSI error variance increases, it still outperforms the reference one regardless of the number of users, even with the highest CSI error level ($\sigma_e^2 = 1$). In particular, the gain achieved by \emph{Prop. Heur.} algorithm even increases with a greater number of users. We can also notice that under all CSI error levels, the proposed algorithm achieves a peak at $K = 60$ users, where the system reaches the maximum level of energy efficiency before slowly decreasing.

\begin{figure}[tbh!]
\centering
\includegraphics[scale=0.23]{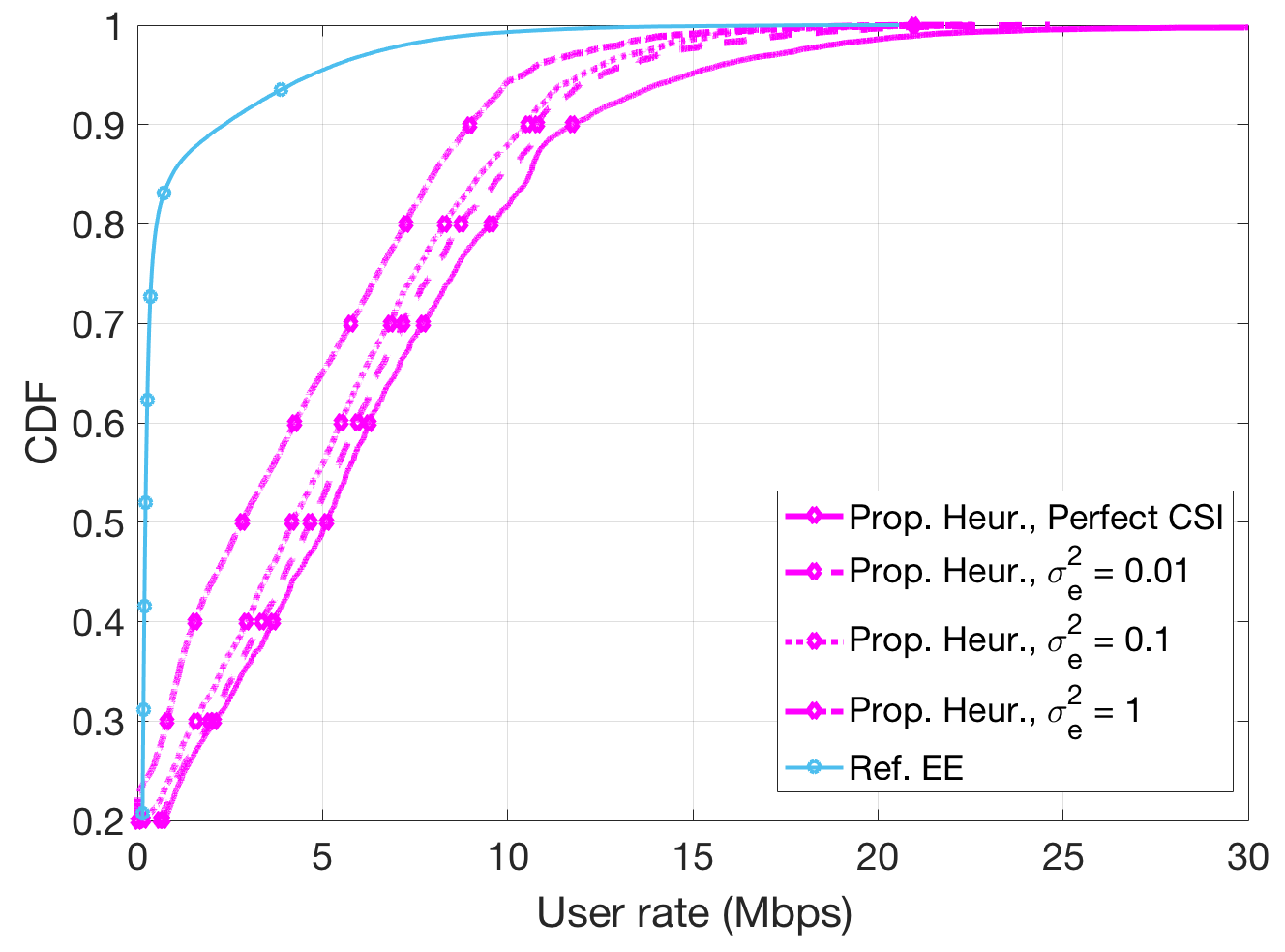}
\caption{CDF of user data rate with various CSI error
levels, proposed and reference algorithms, $K = 60$ users}
\label{fig:CDFUserRate}
\end{figure}

Based on this observation, hereafter, the total number of users is fixed to $K = 60$. Fig. \ref{fig:CDFUserRate} shows the CDF of user data rate given by our proposed heuristic and reference algorithms. Obviously, \emph{Prop. Heur.} achieves higher user rate levels compared to \emph{Ref. EE} for all CSI imperfectness levels.
Moreover, while user data rates also decrease as the CSI error level increases, this degradation appears to be limited. For instance, even with $\sigma_e^2 = 1$, the reduction is limited to 24\% at 80-th percentile, compared to the perfect CSI case. By contrast, \emph{Ref. EE} has a 93\% lower rate compared to \emph{Prop. Heur.} for $\sigma_e^2 = 1$, at the same percentile. In particular, we can observe that the proposed heuristic algorithm increases the proportion of served users in poor channel conditions, including cell edge users, as compared to \emph{Ref. EE}. For instance, the worst 30\% users can achieve up to 1 Mbps under the worst CSI level ($\sigma_e^2 = 1$) and up to 2.3 Mbps for perfect CSI at BBU. By contrast, \emph{Ref. EE} only serves users with better CSI, resulting into 85\% of user data rates below 1 Mbps. Therefore, it can be concluded that our proposed method can clearly provide better user rate-fairness compared to the reference method.

\begin{figure}
\centering
\begin{minipage}{.5\textwidth}
  \centering
  \includegraphics[scale=0.48]{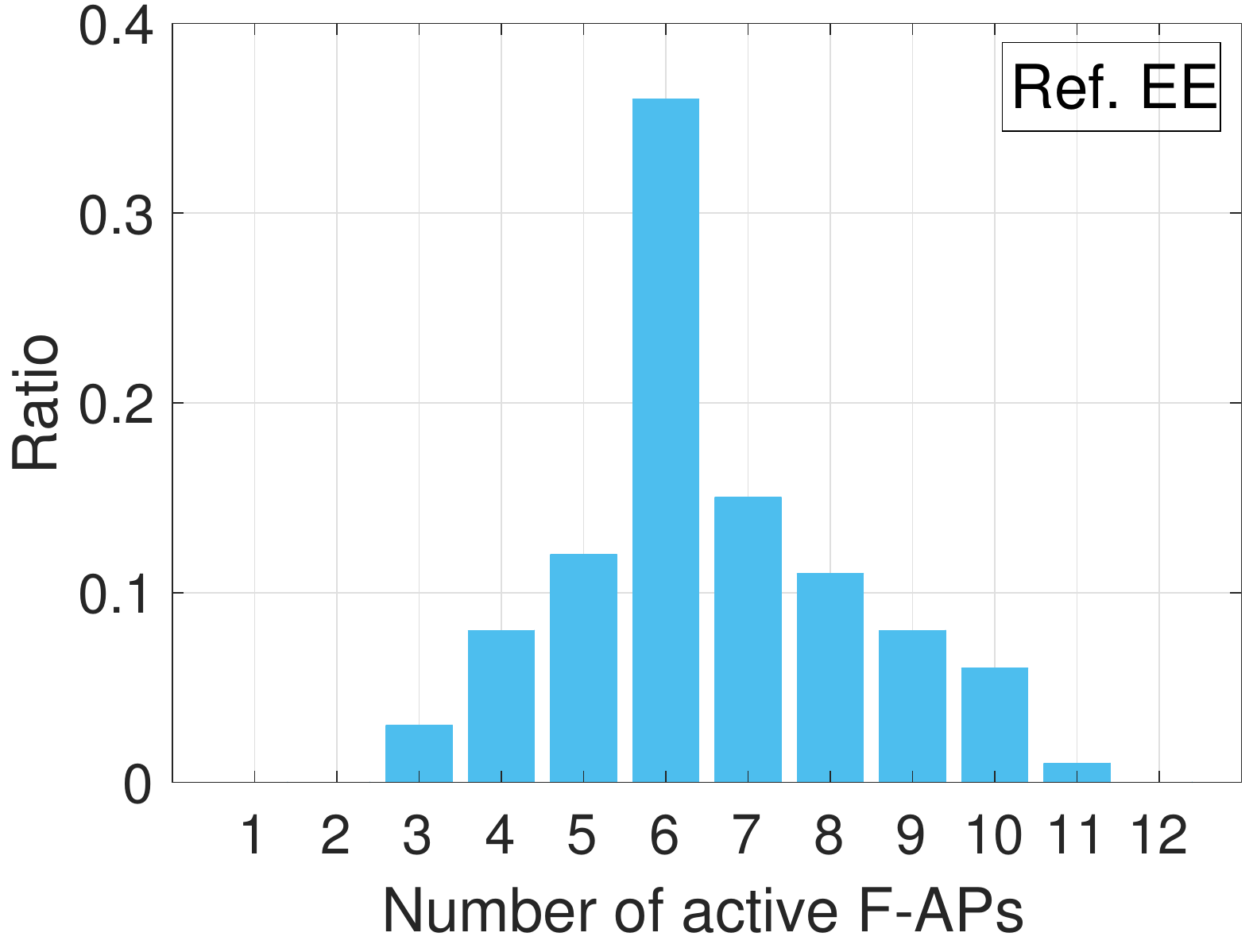}
\end{minipage}%
\begin{minipage}{.5\textwidth}
  \centering
  \includegraphics[scale=0.48]{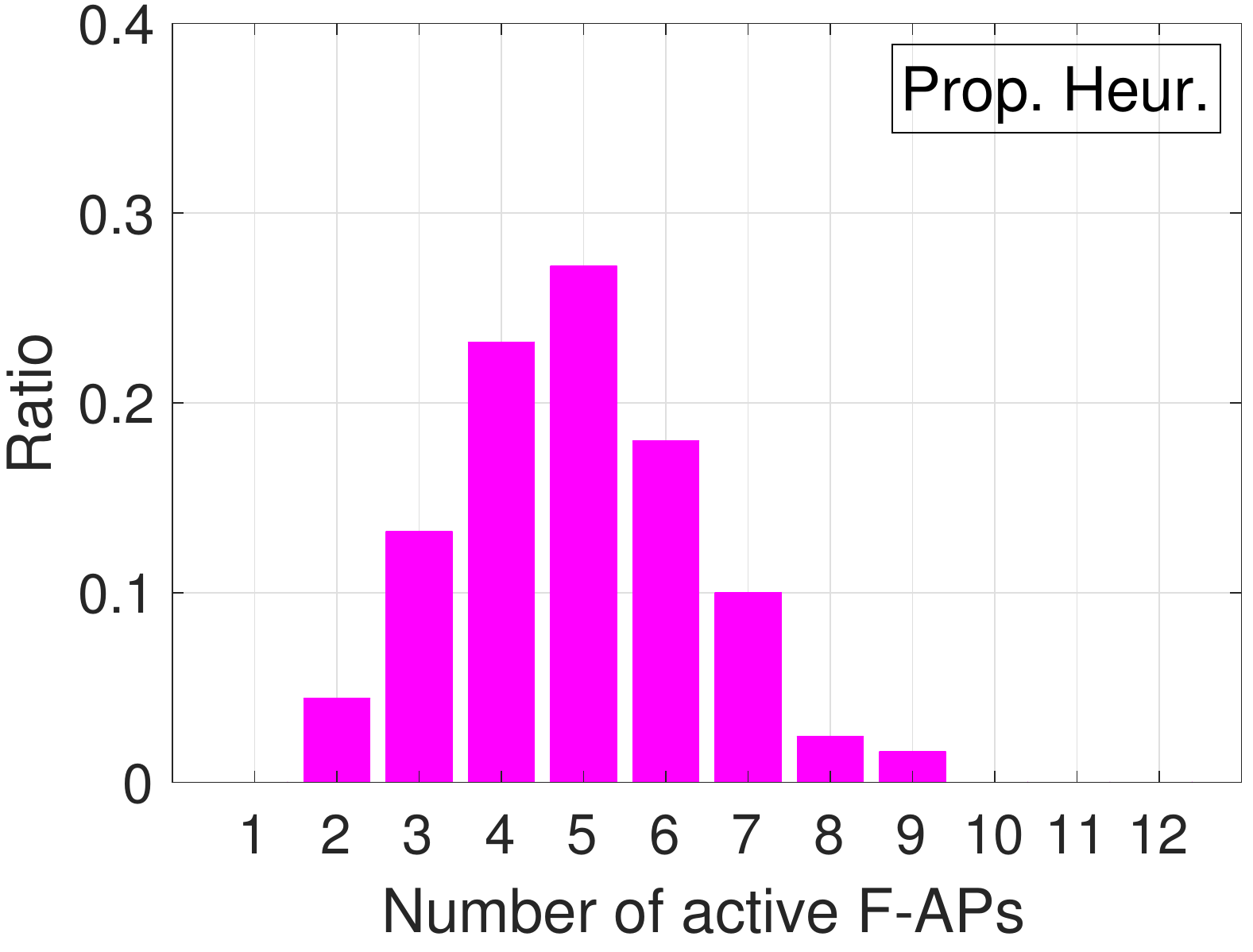}
\end{minipage}
\caption{Distribution of the number of active F-APs for proposed heuristic and reference algorithms, $K = 60$ users}
\label{fig:nbRRHs}
\end{figure}

Furthermore, the distribution of the number of active F-APs is evaluated for the proposed and reference methods in Fig. \ref{fig:nbRRHs}. Again, we observe that the number of active F-APs for \emph{Prop. Heur.} is smaller than that of \emph{Ref. EE}, while providing much higher energy efficiency as well as user rate performance.



Finally, we compare our heuristic algorithm with \emph{Ref. SR} algorithm in \cite{ourpaper,kaneko2019user}, in order to evaluate the trade-off between energy efficiency and sum-rate achieved by \emph{Prop. Heur.} algorithm.

\begin{figure}[h]
\centering
\includegraphics[scale=0.4]{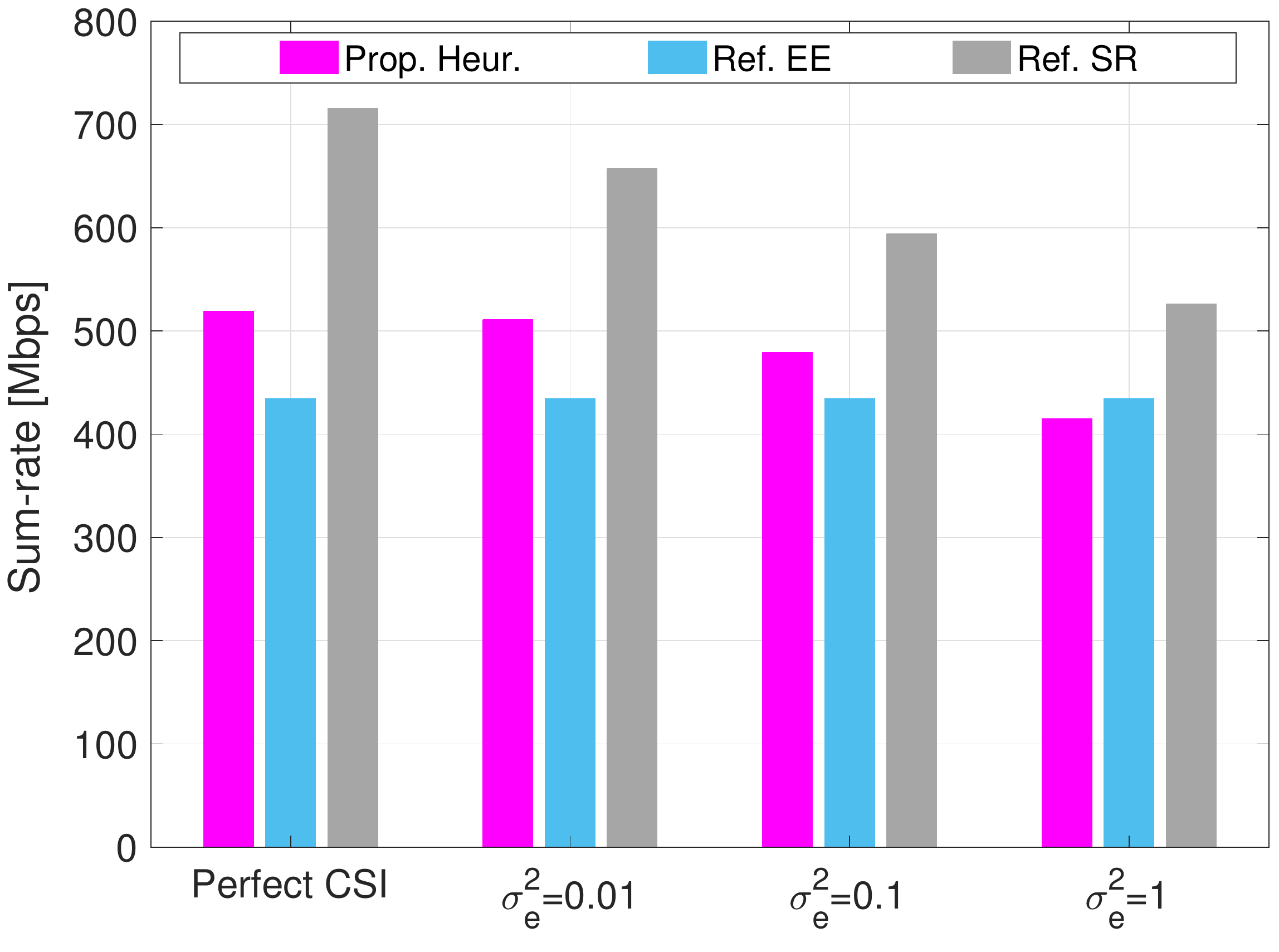}
\caption{Sum-rate performance given by proposed heuristic and reference algorithms, $K = 60$ users}
\label{fig:SRbar}
\end{figure}

Table \ref{tb:SR-EEresults} shows the energy efficiency achieved by both algorithms, \emph{Ref. SR} and \emph{Prop. Heur.}, under the different CSI error variances in the case of 60 users. As expected, \emph{Prop. Heur.} algorithm achieves higher energy efficiency performance for all CSI error levels, whereas the reference scheme achieves higher sum-rate as shown in Fig. \ref{fig:SRbar}. In particular, although the sum-rates offered by \emph{Prop. Heur.} are smaller than those of \emph{Ref. SR}, the sum-rate degradation is limited to around 25\%, while the global energy efficiency of the proposed scheme is considerably higher, i.e., approximately two-fold under all CSI levels. This can be understood as follows. In \emph{Prop. Heur.}, the consumed power is given by (\ref{eq:PF2}), which is obviously lower than that of the reference sum-rate method in (\ref{eq:power}), while still attaining about 75-78\% of the reference sum-rate performance.
\begin{table}[tbh!]
\centering
\caption{Energy efficiency [Mbits/J] given by proposed and reference sum-rate algorithms, $K = 60$ users}
	\setlength{\tabcolsep}{0.5em}	
	{\renewcommand{\arraystretch}{1.1}%
	\begin{tabular*}{0.6\columnwidth}{r|cccc}
		\hline \hline
		&Perfect CSI&$\sigma^2$=0.01&$\sigma^2$=0.1&$\sigma^2$=1 	\\ \hline
		\emph{Prop. Heur.}, $\eta^{Glo}$&38.93&37.00&34.51&29.03\\
		\emph{Ref. SR}, $\eta^{Glo}$&19.0&17.5&15.8&14.0\\ 
		\hline \hline
	\end{tabular*}}
\label{tb:SR-EEresults}
\end{table}

In addition, we can see that \emph{Ref. EE} is not affected by CSI imperfectness, whereas both \emph{Prop. Heur.} and \emph{Ref. SR} offer lower throughput and energy efficiency as the error variance increases. However, the sum-rate given by our heuristic algorithm is still larger than the reference energy-efficiency scheme for all cases except for $\sigma_e^2 = 1$ in which case a slightly lower sum-rate is observed, but where 33\% gains in terms of energy efficiency is attained in counterpart, as observable from Fig. \ref{fig:EEall}. 

%

%% file: conclude.tex
We have investigated the problem of energy efficiency maximization through user association and beamforming in F-RAN,  under the assumption of imperfect CSI knowledge at the cloud BBUs. 
This problem was formulated under the constraints of fronthaul rate limitation, power budget and especially the F-RAN specific requirement ensuring local processing. 
We proposed two approaches, one AL-based and one heuristic method, to handle the mathematical challenges of the formulated optimization problem. In particular, the proposed heuristic strategy utilizes the global but outdated CSI to perform centralized user association and F-AP activation in a periodic manner at the cloud, and the local but perfect CSI for enabling accurate distributed beamforming at each F-AP in every frame. Simulation results proved that, our proposed methods outperformed the reference F-RAN distributed algorithm for a wide range of CSI error levels. Moreover, compared to the baseline sum-rate algorithm, the proposed heuristic scheme could offer notable energy efficiency improvements while limiting the sum-rate degradation.

In the future work, more complex issues of user and device mobility in F-RAN will be considered, as well as IoT application-specific requirements. To tackle such dynamic environments, methods leveraging machine learning techniques will be investigated and developed.